\documentclass[11pt]{JHEP3}
\usepackage{amsmath,amsfonts,slashed}

\newcommand{\bea}{\begin{eqnarray}}
\newcommand{\eea}{\end{eqnarray}}
\newcommand{\be}{\begin{equation}}
\newcommand{\ee}{\end{equation}}
  
\newcommand{\nn}{\nonumber}
\newcommand{\kk}[2]{s_{#1#2}}
\newcommand{\kksq}[2]{s_{#1#2}^2}
\newcommand{\basisb}[1]{A_{#1}}
\newcommand{\basisf}[1]{B_{#1}}
\newcommand{\basisbf}[1]{C_{#1}}

\DeclareMathOperator{\Tr}{Tr}
\DeclareMathOperator{\Sym}{Sym}
\DeclareMathOperator{\Alt}{Alt}

\DeclareMathOperator{\Imag}{Im}

\newdimen\tableauside\tableauside=1ex   
\newdimen\tableaurule\tableaurule=.32pt   
\newdimen\tableaustep
\def\phantomhrule#1{\hbox{\vbox to0pt{\hrule height\tableaurule width#1\vss}}}
\def\phantomvrule#1{\vbox{\hbox to0pt{\vrule width\tableaurule height#1\hss}}}
\def\sqr{\vbox{%
  \phantomhrule\tableaustep
  \hbox{\phantomvrule\tableaustep\kern\tableaustep\phantomvrule\tableaustep}%
  \hbox{\vbox{\phantomhrule\tableauside}\kern-\tableaurule}}}
\def\squares#1{\hbox{\count0=#1\noindent\loop\sqr
  \advance\count0 by-1 \ifnum\count0>0\repeat}}
\def\tableau#1{\vcenter{\offinterlineskip
  \tableaustep=\tableauside\advance\tableaustep by-\tableaurule
  \kern\normallineskip\hbox
    {\kern\normallineskip\vbox
      {\gettableau#1 0 }%
     \kern\normallineskip\kern\tableaurule}%
  \kern\normallineskip\kern\tableaurule}}
\def\gettableau#1 {\ifnum#1=0\let\next=\null\else
  \squares{#1}\let\next=\gettableau\fi\next}

\title{Fermionic superstring loop amplitudes in the \\ pure spinor formalism}

\author{
Christian Stahn\\
Department of Physics, University of North Carolina\\
Chapel Hill, NC 27599--3255, USA\\
E-mail: \email{stahn@physics.unc.edu}}

\abstract{The pure spinor formulation of the ten-dimensional superstring leads to \text{manifestly} supersymmetric loop amplitudes, expressed as integrals in pure spinor superspace.  This paper explores different methods to evaluate these integrals and then uses them to calculate the kinematic factors of the one-loop and two-loop massless four-point amplitudes involving two and four Ramond states.}

\preprint{}

\begin{document}


\section{Introduction}

The quantisation of the ten-dimensional superstring using pure spinors as world-sheet ghosts \cite{bpure} has overcome many difficulties encountered in the Green-Schwarz (GS) and Ramond-Neveu-Schwarz (RNS) formalisms.  Most notably, by maintaining manifest space-time supersymmetry, the pure spinor formalism has yielded super-Poincar\'e covariant multi-loop amplitudes, leading to new insights into perturbative finiteness of superstring theory \cite{bmultiloop,bnewr4}.

Counting fermionic zero modes is a powerful technique in the computation of loop amplitudes in the pure spinor formalism and has for example been used to show that at least four external states are needed for a non-vanishing massless loop amplitude \cite{bmultiloop}.  Furthermore, the structure of massless four-point amplitudes is relatively simple because all fermionic worldsheet variables contribute only through their zero modes.  In the expressions derived for the one-loop \cite{bmultiloop} and two-loop \cite{btwoloop} amplitudes, supersymmetry was kept manifest by expressing the kinematic factors as integrals over pure spinor superspace \cite{bsuperspace} involving three pure spinors~$\lambda$ and five fermionic superspace coordinates~$\theta$,
\be
\label{eq:kinematic}
\begin{split}
K_{\text{1-loop}} &= \bigl\langle (\lambda A) (\lambda\gamma^m W) (\lambda\gamma^n W) {\cal F}_{mn} \bigr\rangle \,, \\
K_{\text{2-loop}} &= \bigl\langle (\lambda \gamma^{mnpqr} \lambda) (\lambda\gamma^s W) {\cal F}_{mn} {\cal F}_{pq} {\cal F}_{rs} \bigr\rangle \,,
\end{split}
\ee
where the pure spinor superspace integration is denoted by $\langle \dots \rangle$, and $A_\alpha(x,\theta)$, $W^\alpha(x,\theta)$ and ${\cal F}_{mn}(x,\theta)$ are the superfields of ten-dimensional Yang-Mills theory. 

The kinematic factors in \eqref{eq:kinematic} have been explicitly evaluated for Neveu-Schwarz states at two loops \cite{bmtwoloop} and one loop \cite{moneloop}, and were found to match the amplitudes derived in the RNS formalism \cite{dhp1}.  This provided important consistency checks in establishing the validity of the pure spinor amplitude prescriptions.  (Related one-loop calculations had been reported in \cite{agv11d}.)

In this paper, it will be shown how to compute the kinematic factors in \eqref{eq:kinematic} when the superfields are allowed to contribute fermionic fields, as is relevant for the scattering of fermionic closed string states as well as Ramond/Ramond bosons.  It turns out that the calculation of fermionic amplitudes presents no additional difficulties, making \eqref{eq:kinematic} a good practical starting point for the computation of four-point loop amplitudes in a unified fashion.  This practical aspect of the supersymmetric pure spinor amplitudes was also emphasised in \cite{ptr4}, where the tree-level amplitudes were used to construct the fermion and Ramond/Ramond form contributions to the four-point effective action of the type II theories.

This paper is organised as follows.  In section 2, different methods to compute pure spinor superspace integrals are explored.  These methods are then applied to the explicit evaluation of the kinematic factors of massless four-point amplitudes at the one-loop level in section 3, and at the two-loop level in section 4.  In both these sections, the bosonic calculations are briefly reviewed before separately considering the cases of two and four Ramond states.  Particular attention will be paid to the constraints imposed by simple exchange symmetries.  An appendix contains algorithms which were used to reduce intermediate expressions encountered in the amplitude calculations to a canonical form.

\section{Zero mode integration}
\label{s:correlators}

The calculation of scattering amplitudes in the pure spinor formalism leads to integrals over zero modes of the fermionic worldsheet variables $\lambda$ and $\theta$.  Both $\theta$ and $\lambda$ are 16-component Weyl spinors, the $\lambda$ are commuting and the $\theta$ anticommuting, and $\lambda$ is subject to the pure spinor constraint $(\lambda\gamma^m\lambda)=0$.  The amplitude prescriptions \cite{bpure,bmultiloop} require three zero modes of $\lambda$ and five zero modes of $\theta$ to be present, and a Lorentz covariant object
\be
\label{eq:tbardef}
{\bar{\cal T}}^{\alpha\beta\gamma,\delta_1\dots\delta_5} \equiv \bigl\langle \lambda^\alpha\lambda^\beta\lambda^\gamma \theta^{\delta_1} \dots \theta^{\delta_5} \bigr\rangle = {\bar{\cal T}}^{(\alpha\beta\gamma),[\delta_1\dots\delta_5]}\\
\ee
was constructed such that the Yang-Mills antighost vertex operator
\be
\label{eq:normalise}
V = (\lambda\gamma^m\theta)(\lambda\gamma^n\theta)(\lambda\gamma^p\theta)(\theta\gamma_{mnp}\theta) \qquad\text{has}\qquad \bigl\langle V \bigr\rangle = 1 \,.
\ee
In this section, different methods of computing such ``pure superspace integrals'' are explored.  As an example, a typical correlator encountered in the two-loop calculations of section 4 is considered:
\be
\label{eq:correx1}
F(k_i, u_i) = k^2_a k^2_m k^3_p k^4_r \bigl\langle (\lambda\gamma^{mnpq[r} \lambda) (\lambda\gamma^{s]} u_1) (\theta\gamma_n{}^{ab} \theta) (\theta\gamma_b u_2) (\theta\gamma_q u_3) (\theta\gamma_s u_4) \bigr\rangle
\ee
Here, $k^i$ and $u_i$ are the momenta and spinor wavefunctions of the four external particles.

\subsection{Symmetry considerations and tensorial formulae}
\label{s:tensorial}

One systematic approach to evaluate the zero mode integrals is to find expressions for all tensors that can be formed from \eqref{eq:tbardef}.  By Fierz transformations, one can always write the product of two $\theta$ spinors as $(\theta\gamma^{[3]}\theta)$, where $\gamma^{[k]}$ denotes the antisymmetrised product of $k$ gamma matrices.  Due to the pure spinor constraint, the only bilinear in $\lambda$ is $(\lambda\gamma^{[5]}\lambda)$, and it is thus sufficient to consider the three cases
\be
\label{eq:gamma135}
\bigl\langle (\lambda\gamma^{[5]}\lambda) (\lambda \{ \gamma^{[1]} \;\text{or}\; \gamma^{[3]} \;\text{or}\; \gamma^{[5]} \} \theta) (\theta\gamma^{[3]}\theta) (\theta\gamma^{[3]}\theta) \bigr\rangle \,.
\ee
Lorentz invariance then implies that it must be possible to express these tensors as sums of suitably symmetrised products of metric tensors, resulting in a parity-even expression, plus a parity-odd part made up from terms which in addition contain an epsilon tensor.  The parity-even parts may be constructed \cite{bmtwoloop} starting from the most general ansatz compatible with the symmetries of the correlator and then using spinor identities along with the normalisation \eqref{eq:normalise} to determine all coefficients in the ansatz.  Duality properties of the spinor bilinears can be used to determine the parity-odd part \cite{moneloop}.  An extensive (and almost exhaustive) list of correlators is found in \cite{bmnonmin}, including the $(\lambda\gamma^{[1]}\theta)$ and $(\lambda\gamma^{[3]}\theta)$ cases of the above list:
\begin{align}
\label{eq:corra8}
\begin{split}
&\bigl\langle (\lambda\gamma^{mnpqr}\lambda) (\lambda\gamma^u\theta) (\theta\gamma^{fgh}\theta) (\theta\gamma^{jkl}\theta) \bigr\rangle \\
& \qquad = -\tfrac{4}{35} \left(\delta^{mnpqr}_{\bar m\bar n\bar p\bar q\bar r} + \tfrac{1}{5!} \varepsilon^{mnpqr}{}_{\bar m\bar n\bar p\bar q\bar r}\right)
\left[ \delta^{\bar m\bar n}_{fg}\delta^{\bar p\bar q}_{jk} (\delta^{\bar r}_l \delta^h_u + \delta^{\bar r}_h \delta^l_u - \delta^{\bar r}_u \delta^h_l ) \right]_{[fgh][jkl]}
\end{split}
\\
\label{eq:corra10}
\begin{split}
&\bigl\langle (\lambda\gamma^{mnpqr}\lambda) (\lambda\gamma^{stu}\theta) (\theta\gamma^{fgh}\theta) (\theta\gamma^{jkl}\theta) \bigr\rangle \\
& \qquad = -\tfrac{24}{35} \left(\delta^{mnpqr}_{\bar m\bar n\bar p\bar q\bar r} + \tfrac{1}{5!} \varepsilon^{mnpqr}{}_{\bar m\bar n\bar p\bar q\bar r}\right)
\left[ \delta^{\bar m}_j \delta^{\bar n\bar p}_{fg} \delta^{\bar q}_{s} \delta^t_l (\delta^{\bar r}_h \delta^k_u - \delta^k_h \delta^{\bar r}_u ) \right]_{[fgh][jkl](fgh\leftrightarrow jkl)}
\end{split}
\end{align}
(Here, the brackets $(fgh\leftrightarrow jkl)$ denote symmetrisation under simultaneous interchange of $fgh$ with $ijk$, with weight one.)  The remaining correlator with the $(\lambda\gamma^{[5]}\theta)$ factor can be derived in the same way, using an ansatz consisting of six parity-even structures.  Taking a trace between the two $\gamma^{[5]}$ factors and noting that
\be
\eta_{ar} \bigl\langle (\lambda\gamma^{mnpqr}\lambda)(\lambda\gamma^{abcde}\theta) \dots \bigr\rangle = -4 \bigl\langle (\lambda\gamma^{mnpq[b}\lambda)(\lambda\gamma^{cde]}\theta) \dots \bigr\rangle \,,
\nn
\ee
one finds a relation to \eqref{eq:corra10}.  This is sufficient to determine all coefficients in the ansatz, and the result is
\begin{multline}
\label{eq:corrnew}
\bigl\langle (\lambda\gamma^{mnpqr}\lambda) (\lambda\gamma^{abcde}\theta) (\theta\gamma^{fgh}\theta) (\theta\gamma^{jkl}\theta) \bigr\rangle = \tfrac{16}{7} \left(\delta^{mnpqr}_{\bar m\bar n\bar p\bar q\bar r} + \tfrac{1}{5!}\varepsilon^{mnpqr}{}_{\bar m\bar n\bar p\bar q\bar r}\right) \\
\times \Bigl[ \delta^{\bar m\bar n\bar p}_{abc} \delta^f_j\delta^{d}_{g}\delta^{\bar q}_{k} ( - \delta^{e}_{h}\delta^{\bar r}_{l} + 2 \delta^{e}_{l}\delta^{\bar r}_{h}) + \delta^{\bar m\bar n}_{ab} \delta^{cd}_{fg}\delta^{\bar p\bar q}_{jk}( \delta^{e}_{h}\delta^{\bar r}_{l} - 3  \delta^{e}_{l}\delta^{\bar r}_{h}) \Bigr]_{[abcde][fgh][jkl](fgh\leftrightarrow jkl)}
\end{multline}

One may find it surprising that the derivation of these tensorial expressions only made use of properties of (pure) spinors, and of the normalisation condition \eqref{eq:normalise}.  However, it can be seen from representation theory that the correlator \eqref{eq:tbardef} is uniquely characterised, up to normalisation, by its symmetry.  To see this, note that \cite{lie} the spinor products $\lambda^3$ and $\theta^5$ transform in
\be
\label{eq:la3th5reps}
\begin{split}
\lambda^{(\alpha}\lambda^\beta\lambda^{\gamma)}  : \qquad \Sym^3 S^+ &= [00003] \oplus [10001] \\
\theta^{[\delta_1} \dots \theta^{\delta_5]}      : \qquad \Alt^5 S^+ &= [00030] \oplus [11010] \,.
\end{split}
\ee
(Here, $\lambda$ and $\theta$ are taken to be in the $S^+$ irrep of SO(1,9), with Dynkin label $[00001]$.)  The tensor product of these contains only one copy of the trivial representation.  This applies to any spinors $\lambda$, which means that the pure spinor property cannot be essential to the derivation of the tensorial identities.  The use of the pure spinor constraint merely allows for simpler derivations of the same identities.

As an illustration of this approach, consider the correlator of eq. \eqref{eq:correx1}.  Leaving the momenta aside for the moment by setting $F=k^2_a k^2_m k^3_p k^4_r\tilde{F}$, the task is to compute
\be
\tilde{F} = \bigl\langle (\lambda\gamma^{mnpq[r} \lambda) (\lambda\gamma^{s]} u_1) (\theta\gamma_n{}^{ab} \theta) (\theta\gamma_b u_2) (\theta\gamma_q u_3) (\theta\gamma_s u_4) \bigr\rangle \,.
\nn
\ee
After applying two Fierz transformations,
\bea
\tilde{F} &=& \Bigl( \phantom{++} \tfrac{1}{16} \; \bigl\langle (\lambda\gamma^{mnpq[r|}\lambda) (\lambda\gamma^{c}\theta) (\theta\gamma_n{}^{ab}\theta)(\theta\gamma^{jkl}\theta) \bigr\rangle \; (u_1\gamma^{|s]}\gamma_{c}\gamma_b u_2)
\nn \\
 && \phantom{\Bigl(} + \tfrac{1}{3!\cdot 16} \; \bigl\langle (\lambda\gamma^{mnpq[r|}\lambda)(\lambda\gamma^{cde}\theta) (\theta\gamma_n{}^{ab}\theta)(\theta\gamma^{jkl}\theta) \bigr\rangle \; (u_1\gamma^{|s]}\gamma_{cde}\gamma_b u_2)
\nn \\
 && \phantom{\Bigl(} + \tfrac{1}{2\cdot 5!\cdot 16} \; \bigl\langle (\lambda\gamma^{mnpq[r|}\lambda)(\lambda\gamma^{cdefg}\theta) (\theta\gamma_n{}^{ab}\theta)(\theta\gamma^{jkl}\theta) \bigr\rangle \; (u_1\gamma^{|s]}\gamma_{cdefg}\gamma_b u_2) \Bigr) 
\nn \\
&& \times \tfrac{1}{3!\cdot 16} (u_3\gamma_q\gamma_{jkl}\gamma_s u_4) \,,
\nn
\eea
one obtains a combination of the fundamental correlators listed in \eqref{eq:corra8}, \eqref{eq:corra10} and \eqref{eq:corrnew}.  A~reliable evaluation of the numerous index symmetrisations is made possible by the use of a computer algebra program.  In doing these calculations with Mathematica, an essential tool is the GAMMA package \cite{gamma}, expanding the products of gamma matrices in a $\gamma^{[k]}$~basis.  The result consists of two parts, $\tilde{F}=\tilde{F}^{(\delta)} + \tilde{F}^{(\varepsilon)}$, with
\bea
\tilde{F}^{(\delta)} &=& \tfrac{1}{560} (u_1\gamma^{mpr} u_2)(u_3\gamma^a u_4) + \tfrac{7}{720} \delta^a_p\delta^m_r (u_1\gamma^i u_2)(u_3\gamma_i u_4)  + \dots \nn \\
\label{eq:ftildedelta}
&& - \tfrac{1}{1680} (u_1\gamma^{ai_1i_2} u_2)(u_3\gamma^{mpr}{}_{i_1i_2} u_4) \qquad\text{(92 terms)} \\
\tilde{F}^{(\varepsilon)} &=& -\tfrac{1}{1209600} \varepsilon_{i_1\dots i_7}{}^{mpr} (u_1\gamma^{i_1\dots i_7} u_2)(u_3\gamma^a u_4) + \dots 
\nn \\
&& - \tfrac{1}{604800} \varepsilon^{ampr}{}_{i_1\dots i_6} (u_1\gamma^{i_3\dots i_9} u_2)(u_3\gamma^{i_1i_2}{}_{i_7i_8i_9} u_4) \qquad\text{(34 terms)}
\label{eq:ftildeeps1}
\eea
The epsilon tensors in the second part can be eliminated using the fact that the $u_i$ are chiral spinors: If all the indices on $\gamma^{[k]} u_i$ are contracted into an epsilon tensor, one uses
\be
\label{eq:epsgamma1}
\varepsilon_{i_1\dots i_{k'} j_1\dots j_k} \gamma^{j_1\dots j_k} \gamma^{11} = (-)^{\tfrac{1}{2}k(k+1)} k! \; \gamma_{i_1\dots i_{k'}} \,,
\ee
where $\gamma^{11} = \tfrac{1}{10!}\varepsilon_{i_0\dots i_9} \gamma^{i_0\dots i_9}$.  More generally, if all but $r$ indices of $\gamma^{[k]} u_i$ are contracted,
\be
\label{eq:epsgamma2}
\varepsilon_{i_1\dots i_{k'} j_1\dots j_k} \gamma^{p_1\dots p_r j_1\dots j_k} \gamma^{11} =  (-)^{\tfrac{1}{2}k(k+1)} k! \frac{k'!}{(k'-r)!} \; \delta^{p_r\dots p_1}_{[i_1\dots i_r} \gamma_{i_{r+1}\dots i_k']} \,.
\ee
The result of these manipulations is
\begin{align}
\tilde{F}^{(\varepsilon)} =& - \tfrac{1}{560} (u_1\gamma^{mpr} u_2)(u_3\gamma^a u_4) - \tfrac{1}{280} \delta^p_r (u_1\gamma^{ami} u_2)(u_3\gamma_{i} u_4) + \dots \nn \\
\label{eq:ftildeeps2}
& + \tfrac{9}{11200} (u_1\gamma^{i_1i_2i_3} u_2)(u_3\gamma^{ampr}{}_{i_1i_2i_3} u_4) \qquad\text{(53 terms)}
\end{align}
(Note that while the epsilon terms in the basic correlator formulae were easily obtained from the delta terms by using Poincar\'e duality, this cannot be done here in any obvious way.)  The last step in the evaluation of \eqref{eq:correx1} is to contract with the momenta, $F=k^2_a k^2_m k^3_p k^4_r\tilde{F}$, and to simplify the expressions using the on-shell identities $\sum_i k_i=0$, $k_i^2=0$, $\slashed{k}_i u_i = 0$.  It is shown in appendix \ref{s:basis4f} that there are only ten independent scalars, denoted by $\basisf{1}\dots\basisf{10}$, that can be formed from four momenta and the four spinors $u_1\dots u_4$.  With respect to this basis, the result is
\bea
F^{(\delta)} &=& \tfrac{1}{48\cdot 10080} \left( 695 \kk{1}{2} (u_1 \slashed{k}_3 u_2)(u_3 \slashed{k}_1 u_4) + \dots + 233 \kksq{1}{3} (u_1 \gamma^a u_2)(u_3 \gamma_a u_4) \right) \qquad\text{(7 terms)} \nn\\
 &=& \tfrac{1}{48\cdot 10080}(695,775,0,-80,356,356,0,233,233,0)_{\basisf{1}\dots\basisf{10}} \,, \nn \\
F^{(\varepsilon)} &=& \tfrac{1}{48\cdot 10080}(-23,-7,0,-16,28,28,0,7,7,0)_{\basisf{1}\dots\basisf{10}} \,, \nn \\
\label{eq:ftotal}
F &=& \tfrac{1}{10080}(14,16,0,-2,8,8,0,5,5,0)_{\basisf{1}\dots\basisf{10}} \,,
\eea
where $\kk{i}{j}=k_i\cdot k_j$.

\subsection{A spinorial formula}
\label{s:tracemethod}

While the derivation of tensorial identities for correlators of the form \eqref{eq:gamma135} is relatively straightforward and elegant, it may be a tedious task to transform the expressions encountered in amplitude calculations to match this pattern.  As seen in the example calculated above, this is particularly true if additional spinors are involved, making it necessary to apply Fierz transformations.  It is therefore desirable to use a covariant correlator expression with open spinor indices.  Such an expression was given in \cite{bpure,bmultiloop}:
\be
\label{eq:corrformula}
\bar{T}^{\alpha\beta\gamma,\delta_1\dots\delta_5} = N^{-1} \left[ (\gamma^m)^{\alpha\delta_1} (\gamma^n)^{\beta\delta_2} (\gamma^p)^{\gamma\delta_3} (\gamma_{mnp})^{\delta_4\delta_5} \right]_{(\alpha\beta\gamma)[\delta_1\dots\delta_5]} \,,
\ee
where $N$ is a normalisation constant and the brackets $()[]$ denote (anti-)symmetrisation with weight one.  (Note that the right hand side is automatically gamma-matrix traceless: any gamma-trace
\be
(\gamma^r)_{\alpha\beta} \times (\gamma^m)^{\alpha [\delta_1|} (\gamma^n)^{\beta |\delta_2|} (\gamma^p)^{\gamma|\delta_3} (\gamma_{mnp})^{\delta_4\delta_5]} = - (\gamma^{mnr})^{[\delta_1\delta_2} (\gamma_{mnp})^{\delta_3\delta_4} (\gamma^p)^{\delta_5]\gamma} = 0
\nn
\ee
vanishes due to the double-trace identity $(\gamma_{ab}\theta)^\alpha (\theta\gamma^{abc}\theta)=0$, which follows from the fact that the tensor product $(\Alt^3 S^+)\otimes S^-$ does not contain a vector representation and therefore the vector $(\psi\gamma_{ab}\theta)(\theta\gamma^{abc}\theta)$ has to vanish for all spinors $\psi$, and can also be shown by applying a Fierz transformation.)  This prescription was originally motivated \cite{bmultiloop} by the fermionic expansion of the Yang-Mills antighost vertex operator $V$,
\be
\label{eq:tdef}
V = T_{\alpha\beta\gamma,\delta_1\dots\delta_5} \lambda^\alpha\lambda^\beta\lambda^\gamma \theta^{\delta_1}\dots \theta^{\delta_5}
\ee
with
\be
T_{\alpha\beta\gamma,\delta_1\dots\delta_5} = \bigl[ (\gamma^m)_{\alpha\delta_1} (\gamma^n)_{\beta\delta_2} (\gamma^p)_{\gamma\delta_3} (\gamma_{mnp})_{\delta_4\delta_5} \bigr]_{(\alpha\beta\gamma)[\delta_1\dots\delta_5]} \,,
\nn
\ee
where $T$ is related to $\bar{T}$ by a parity transformation, up to the overall constant $N$.  (Since $\bar{T}$ is uniquely determined by its symmetries, any covariant expression will be proportional to $\bar{T}$, after symmetrisation of the spinor indices, and this is merely the simplest choice.)

Equation \eqref{eq:corrformula} immediately yields an algorithm to convert any correlator into traces of gamma matrices or, if additional spinors are involved, bilinears in those spinors.  It is, however, already very tiresome to determine the normalisation constant $N$ by hand.  The main advantage of this approach is that it clearly lends itself to implementation on a computer algebra system, which can easily carry out the spinor index symmetrisations, simplify the gamma products (again using the GAMMA package), and compute the traces.  For example,
\bea
N \langle V \rangle &=& \left[ (\gamma^m)^{\alpha\delta_1} (\gamma^n)^{\beta\delta_2} (\gamma^p)^{\gamma\delta_3} (\gamma_{mnp})^{\delta_4\delta_5} \right]_{(\alpha\beta\gamma)[\delta_1\dots\delta_5]} (\gamma_x)_{\alpha\delta_1} (\gamma_y)_{\beta\delta_2} (\gamma_z)_{\gamma\delta_3} (\gamma^{xyz})_{\delta_4\delta_5} \nn \\
&=& -\tfrac{1}{60} \Tr(\gamma_x\gamma^m) \Tr(\gamma_y\gamma^n) \Tr(\gamma_z\gamma^p) \Tr(\gamma^{xyz}\gamma_{pnm}) + \dots \nn \\
&& - \tfrac{1}{60} \Tr(\gamma_z\gamma_{pnm} \gamma^{zyx} \gamma^n \gamma_x \gamma^m \gamma_y \gamma^p) \hfill \qquad\text{(60 terms)} \nn \\
&=& 5160960 \,.
\nn
\eea
The correct normalisation is therefore obtained by setting $N=5160960$.

Returning to the example correlator \eqref{eq:correx1}, one finds that the calculation is by far simpler than with the previous method.  After carrying out the symmetrisations $(\alpha\beta\gamma)[\delta_i]$, one obtains
\bea
N\tilde{F} &=& \tfrac{1}{60} \Tr(\gamma_x\gamma^{ab}{}_n\gamma_y\gamma^{mnpq[r|}) (u_3\gamma_q\gamma^{xyz}\gamma_s u_4) (u_1\gamma^{|s]}\gamma_z\gamma_b u_2) + \dots \nn\\
&& - \tfrac{1}{30} (u_2 \gamma_b \gamma^{xyz} \gamma_q u_3)(u_1 \gamma_s\gamma_y\gamma^{ab}{}_n\gamma_x\gamma^{mnpq[r}\gamma_z\gamma^{s]} u_4) \,, \qquad\text{(24 terms)}
\nn
\eea
where elementary index re-sorting has reduced the number of terms from 60 to 24.  Expanding the gamma products leads to
\be
N \tilde{F} = \tfrac{476}{5} \delta^p_r (u_1\gamma^m u_4)(u_2\gamma^a u_3) + \dots + \tfrac{8}{15} (u_1\gamma^{ai_1i_2i_3i_4} u_2)(u_3\gamma^{mpr}{}_{i_1i_2i_3i_4} u_4) \,, \qquad\text{(294 terms)}
\nn
\ee
which, in contrast to \eqref{eq:ftildeeps1}, contains no epsilon terms as there are not enough free indices present.  Note that this intermediate result contains terms with with $u_1$ paired with $u_3$ or $u_4$, so it is not possible to directly compare to eqs. \eqref{eq:ftildedelta} and \eqref{eq:ftildeeps2}.  However, after contracting with the momenta $k^2_ak^2_mk^3_pk^4_r$ and decomposing the result in the basis $\basisf{1}\dots\basisf{10}$, one again obtains
\be
F = \tfrac{1}{10080}(14,16,0,-2,8,8,0,5,5,0)_{\basisf{1}\dots\basisf{10}} \,,
\ee
in agreement with \eqref{eq:ftotal}.

The algorithm just outlined will be the method of choice for all correlator calculations in the later sections of this paper and can easily be applied to a wider range of problems.  The only limitation is that the larger the number of gamma matrices and open indices of the correlator, the slower the computer evaluation will be.  For example, the correlator considered in eq. (5.2) of \cite{bmnonmin},
\begin{multline}
\label{eq:t10}
t^{mnm_1n_1\dots m_4n_4}_{10} \equiv \bigl\langle (\lambda\gamma^p\gamma^{m_1n_1}\theta) (\lambda\gamma^q\gamma^{m_2n_2}\theta) (\lambda\gamma^r\gamma^{m_3n_3}\theta)  (\theta\gamma^{m}\gamma^{n}\gamma_{pqr}\gamma^{m_4n_4}\theta) \bigr\rangle \\
= -\tfrac{2}{45} \left( \eta^{mn} t_8^{m_1n_1\dots m_4n_4} - \tfrac{1}{2} \varepsilon^{mnm_1n_1\dots m_4n_4} \right) \,,
\end{multline}
can still be verified with this method but this already requires substantial runtime.

\subsection{Component-based approach}

A third method to evaluate the zero mode integrals consists of choosing a gamma matrix representation, expanding the integrand as a polynomial in spinor components, and then applying \eqref{eq:corrformula} to the individual monomials.  This procedure seems particularly appealing if at some stage of the calculation one works with a matrix representation anyhow, in order to reduce the results to a canonical form (e.g. as outlined in appendix \ref{s:spinors}).  An efficient decomposition algorithm (of $k^4 u_1u_2u_3u_4$ scalars, say) only needs a few non-zero momentum and spinor wavefunction components to distinguish all independent scalars, and therefore $k$ and $u$ can be replaced by sparse vectors.  Furthermore, a trivial observation allows for a much quicker numeric evaluation of correlator components than a naive use of \eqref{eq:corrformula}:  In view of \eqref{eq:tdef}, one can equivalently compute the components of the parity-transformed expression $\bar{V} = (\bar{\lambda}\gamma^m\bar{\theta}) (\bar\lambda\gamma^n\bar\theta) (\bar\lambda\gamma^p\bar\theta) (\bar\theta\gamma_{mnp}\bar\theta)$, where $\bar\lambda$ and $\bar\theta$ are spinors of chirality opposite to that of $\lambda$, $\theta$.  In the representation given in appendix \ref{s:gammamat}, $\bar{V}$ coincides with $V|_{\lambda \to \bar\lambda, \theta \to \bar\theta}$, and
\be
V = 
192 \, \lambda^9\lambda^9\lambda^9 \theta^1\theta^2\theta^3\theta^4\theta^9 + \dots + 480 \, \lambda^1\lambda^2\lambda^3
\theta^1\theta^9\theta^{10}\theta^{13}\theta^{15} + \dots \qquad\text{(100352 terms)}
\nn
\ee
The monomials in the fermionic expansion of $\bar{V}$ then correspond to the arguments of non-zero correlators, and the coefficients of those monomials are, up to normalisation and symmetry factors, the correlator values.

Unfortunately, it turns out that the complexity of typical correlators (e.g. the one given in \eqref{eq:correx1}) makes it difficult to carry out the expansion in fermionic components in any straightforward way and limits this method to special applications.  For example, the coefficients in \eqref{eq:t10} can be checked relatively easily by choosing particular index values, such as
\begin{multline}
\bigl\langle (\lambda\gamma^p\gamma^{12}\theta) (\lambda\gamma^q\gamma^{21}\theta) (\lambda\gamma^r\gamma^{34}\theta)  (\theta\gamma^{0}\gamma^{0}\gamma_{pqr}\gamma^{43}\theta) \bigr\rangle \\
= \bigl\langle 12 \, \lambda^1\lambda^1\lambda^1 \theta^1\theta^9\theta^{10}\theta^{11}\theta^{12} + \dots + 12 \, \lambda^{16}\lambda^{16}\lambda^{16} \theta^5\theta^6\theta^7\theta^8\theta^{16} \bigr\rangle 
= \tfrac{1}{45} \,.
\nn
\end{multline}
(For fixed values of $pqr$, one gets no more than about $10^5$ monomials of the form $\lambda^3 \theta^5$).  This approach may thus still be helpful in situations where the result has been narrowed down to a simple ansatz.

\section{One-loop amplitudes}
\label{s:oneloop}

The amplitude for the scattering of four massless states of the type IIB superstring was computed \cite{bmultiloop} in the pure spinor formalism as
\be
{\cal A} = K \bar{K} \int \frac{d^2 \tau}{(\Imag \tau)^5} \int d^2 z_2 \int d^2 z_3 \int d^2 z_4 \prod_{i<j} G(z_i, z_j)^{k_i\cdot k_j} \,,
\ee
where $G(z_i, z_j)$ is the scalar Green's function, and the kinematic factor is given by the product $K\bar{K}$ of left- and right-moving open superstring expressions,
\be
\label{eq:koneloop}
K_{\text{1-loop}} = \bigl\langle (\lambda A_1) (\lambda\gamma^{m} W_2)(\lambda\gamma^{n}W_3){\cal F}_{4,mn} \bigr\rangle + \bigl(\text{cycl(234)}\bigr) \,.
\ee
Here the indices $1\dots 4$ label the external states and ``$\dotsb + \bigl(\text{cycl(234)}\bigr)$'' denotes the addition of two other terms obtained by cyclic permutation of the indices $234$.  The spinor superfield $A_\alpha$ and its supercovariant derivatives, the vector gauge superfield $A_m = \tfrac{1}{8} \gamma_m^{\alpha\beta} D_\alpha A_\beta$ as well as the spinor and vector field strengths $W^{\alpha} = \tfrac{1}{10} (\gamma^m)^{\alpha\beta} (D_\beta A_m - \partial_m A_{\beta})$ and ${\cal F}_{mn} = \tfrac{1}{8} (\gamma_{mn})^\alpha{}_\beta D_\alpha W^\beta = 2 \partial_{[m} A_{n]}$, describe ten-dimensional super-Yang-Mills theory.  The physical fields of this theory, a gauge boson and a gaugino, are found in the leading components $A_m| = \zeta_m$ and $W^\alpha| = \hat{u}^\alpha$ and correspond to the Neveu-Schwarz and Ramond superstring states.

The superfields $A_\alpha$ and $W^\alpha$ as well as the gaugino field $\hat{u}^\alpha$ are anticommuting.\footnote{Thanks to Carlos Mafra for pointing this out.}  To facilitate computer calculations involving polynomials in the spinor components, and for easier comparison with the literature, it will be more convenient to work with commuting fermion wavefunctions $u^\alpha$.  Fortunately, as the kinematic factors with fermionic external states are multilinear functions of the distinctly labelled spinors $\hat{u}_i$, it is straightforward to translate between the two conventions: Any monomial expression in $\hat{u}_1 \dots \hat{u}_4$ (and possibly fermionic coordinates $\theta$) corresponds to the same expression in $u_1 \dots u_4$, multiplied by the signature of the permutation sorting the $\hat{u}_i$ (and any $\theta$ variables) into some fixed order, such as $(\theta\cdots\theta) \hat{u}_1^{\alpha_1}\hat{u}_2^{\alpha_2}\hat{u}_3^{\alpha_3}\hat{u}_4^{\alpha_4}$.

Choosing a gauge where $\theta^\alpha A_\alpha = 0$, the on-shell identities
\be
2 D_{(\alpha} A_{\beta)} = \gamma^m_{\alpha\beta} A_m \,,\qquad D_\alpha W^\beta = \tfrac{1}{4} (\gamma^{mn})_\alpha{}^\beta {\cal F}_{mn}
\nn
\ee
have been used to derive recursive relations \cite{ptr4,holography,vertex} for the fermionic expansion
\be
A_{\alpha}^{(n)} = \tfrac{1}{n+1} (\gamma^m\theta)_{\alpha} A_m^{(n-1)} \,,\qquad
A_m^{(n)} = \tfrac{1}{n} (\theta\gamma_m W^{(n-1)}) \,,\qquad
W^{\alpha(n)} = - \tfrac{1}{2n} (\gamma^{mn}\theta)^\alpha \partial_m A_n^{(n-1)} \,,
\nn
\ee
where $f^{(n)} = \tfrac{1}{n!} \theta^{\alpha_n}\cdots\theta^{\alpha_1} (D_{\alpha_1}\cdots D_{\alpha_n} f)|$.  These recursion relations were explicitly solved in \cite{ptr4}, reducing the fermionic expansion to a simple repeated application of the derivative operator ${\cal O}_m{}^q = \tfrac{1}{2} (\theta\gamma_m{}^{qp}\theta)\partial_p$:
\be
\label{eq:thetaex}
\begin{split}
A_{m}^{(2k)} &= \tfrac{1}{(2k)!} [ {\cal O}^k ]_m{}^q \zeta_q \,, \\
A_{m}^{(2k+1)} &= \tfrac{1}{(2k+1)!} [ {\cal O}^k ]_m{}^q (\theta\gamma_q \hat{u}) \,.
\end{split}
\ee
With this solution at hand, one has all ingredients to evaluate the kinematic factor \eqref{eq:koneloop} for the three cases of zero, two, or four fermionic states.

\subsection{Review: four bosons}
\label{s:oneloop4b}

The kinematic factor involving four bosons was considered in \cite{moneloop} and this calculation will now be reviewed briefly.  First, note that the outcome is not fixed by symmetry: The result must be gauge invariant \cite{bmultiloop} and therefore expressible in terms of the field strengths $F_1\dots F_4$.  The cyclic symmetrisation in \eqref{eq:koneloop} yields expressions symmetric in $F_2$, $F_3$, $F_4$, and acting on scalars constructed from the $F_i$ only, the $(234)$ symmetrisation is equivalent to complete symmetrisation in all labels (1234).  Thus the result must be a linear combination of the two gauge invariant symmetric $F^4$ scalars, namely the single trace $\Tr(F_{(1}F_{2}F_{3}F_{4)})$ and double trace $\Tr(F_{(1}F_{2}) \Tr(F_{3}F_{4)})$, leaving one relative coefficient to be determined.

Since all four states are of the same kind, one may first evaluate the correlator for one labelling and then carry out the cyclic symmetrisation:
\be
K^{\text{(4B)}}_{\text{1-loop}} = \left. \bigl\langle (\lambda A_1) (\lambda\gamma^{m} W_2)(\lambda\gamma^{n}W_3){\cal F}_{4,mn} \bigr\rangle \right|_{\text{4B}} + \bigl(\text{cycl (234)}\bigr) \,.
\nn
\ee
The different ways to saturate $\theta^5$ result in a sum of terms of the form
\be
\label{eq:xabcd}
X_{ABCD} = \left\langle (\lambda A^{(A)}_1) (\lambda\gamma^{m } W^{(B)}_2)(\lambda\gamma^{n}W^{(C)}_3){\cal F}^{(D)}_{4,mn} \right\rangle
\ee
with $A+B+C+D=5$ and $A$, $B$, $C$ odd, $D$ even:
\be
\left. \bigl\langle (\lambda A_1) (\lambda\gamma^{m} W_2)(\lambda\gamma^{n}W_3){\cal F}_{4,mn} \bigr\rangle \right|_{\text{4B}}
= X_{3110} + X_{1310} + X_{1130} + X_{1112} \,.
\nn
\ee
Note that $X_{1310}$ and $X_{1130}$ are related by exchange of the labels 2 and 3.  This exchange can be carried out after computing the correlator, an operation which will in the following be denoted by $\pi_{23}$.  Using \eqref{eq:thetaex} for the superfield expansions and replacing $\partial_m\to ik_m$, one obtains
\bea
X_{3110} &= -\tfrac{1}{512} F^1_{mn} F^2_{pq} F^3_{rs} F^4_{tu} {\tilde X}_{3110} \,,\;\;
&{\tilde X}_{3110} = \bigl\langle (\lambda\gamma^{[t|}\gamma^{pq}\theta) (\lambda\gamma^{|u]}\gamma^{rs}\theta) (\lambda\gamma_a\theta)(\theta\gamma^{amn}\theta) \bigr\rangle \,,
\nn \\
X_{1112} &= -\tfrac{1}{128} i k^4_m \zeta^1_n F^2_{pq} F^3_{rs} F^4_{tu} {\tilde X}_{1112} \,,\;\;
&{\tilde X}_{1112} = \bigl\langle (\lambda\gamma^{[m|}\gamma^{pq}\theta) (\lambda\gamma^{|a]}\gamma^{rs}\theta) (\lambda\gamma^n\theta)(\theta\gamma_a{}^{tu}\theta) \bigr\rangle \,,
\nn \\
X_{1310} &= -\tfrac{1}{384} i k^3_m \zeta^1_n F^2_{pq} F^3_{rs} F^4_{tu} {\tilde X}_{1310} \,,\;\;
&{\tilde X}_{1310} = \bigl\langle (\lambda\gamma^{[t|}\gamma^{ma}\theta) (\lambda\gamma^{|u]}\gamma^{rs}\theta) (\lambda\gamma^n\theta)(\theta\gamma_a{}^{pq}\theta) \bigr\rangle \,.
\nn
\eea
The method outlined in section \ref{s:tracemethod} is readily applicable to these correlators.  For example, for $X_{3111}$, the trace evaluation yields
\bea
{\tilde X}_{3110} &=& N^{-1} \Bigl[ \tfrac{1}{60} \Tr(\gamma_a\gamma^z) \Tr(\gamma_{xyz}\gamma^{anm}) \Tr(\gamma^x\gamma_{qp}\gamma^{[t|}) \Tr(\gamma^y\gamma_{sr}\gamma^{|u]}) + \dotsb \nn \\ 
&& \dotsb +\tfrac{1}{60} \Tr(\gamma^{[u|} \gamma_{rs} \gamma_{zyx} \gamma_{qp} \gamma^{|t]} \gamma^x \gamma_a \gamma^y \gamma^{mna} \gamma^z ) \Bigr] \qquad\text{(60 terms)}
\nn \\
 &=& \left(\tfrac{2}{35} \delta^{mp}_{rs}\delta^{nq}_{tu} - \tfrac{1}{315} \delta^{mn}_{tu}\delta^{pq}_{rs} - \tfrac{1}{45} \delta^{mn}_{rs}\delta^{pq}_{tu} + \tfrac{26}{315} \delta^{mn}_{pr}\delta^{qs}_{tu} \right)_{[mn][pq][rs][tu](pq\leftrightarrow rs)}
\nn
\eea
Upon contracting with the field strengths, momenta and polarisations, and symmetrising over the cyclic permutations (234) (with weight 3), one finds that all three contributions are separately gauge invariant:
\bea
X_{3110} + \bigl(\text{cycl(234)}\bigr) &=& -\tfrac{11}{13440}\Tr(F_{(1}F_2F_3F_{4)}) + \tfrac{1}{6720} \Tr(F_{(1}F_2) \Tr(F_3 F_{4)}) \nn\\
X_{1112} + \bigl(\text{cycl(234)}\bigr) &=& -\tfrac{19}{53760} \Tr(F_{(1}F_2F_3F_{4)}) + \tfrac{31}{215040} \Tr(F_{(1}F_2) \Tr(F_3 F_{4)}) \nn\\
(1+\pi_{23})X_{1310} + \bigl(\text{cycl(234)}\bigr) &=& -\tfrac{1}{10240} \left( 4 \Tr(F_{(1}F_2F_3F_{4)}) - \Tr(F_{(1}F_2) \Tr(F_3 F_{4)}) \right) \nn
\eea
The sum $X_{3110} + X_{1112}$ has the right ratio of single- and double-trace terms to be proportional to the well-known result $t_8F^4$, and the last line exhibits the right ratio by itself.
The overall kinematic factor is therefore
\be
K^{\text{4B}}_{\text{1-loop}} = -\tfrac{1}{2560} \left( 4 \Tr(F_{(1}F_2F_3F_{4)}) - \Tr(F_{(1}F_2) \Tr(F_3 F_{4)}) \right) = -\tfrac{1}{15360} t_8 F^4 \,,
\ee
in agreement with the expressions derived in the RNS \cite{aticksen} and Green-Schwarz \cite{gsloop} formalisms.

\subsection{Four fermions}
\label{s:oneloop4f}

The four-fermion kinematic factor could be evaluated in the same way as in the four-boson case by summing up all terms $X_{ABCD}$, $A+B+C+D=5$, now with $A$, $B$, $C$ even and $D$ odd.  Note however that this time, the outcome is fixed by symmetry:  The cyclic symmetrisation in~\eqref{eq:koneloop} leads to a completely symmetric dependence on $\hat{u}_2$, $\hat{u}_3$, $\hat{u}_4$, and therefore to a completely antisymmetric dependence on $u_2$, $u_3$, $u_4$.  Acting on scalars of the form $k^2 u_1u_2u_3u_4$, antisymmetrising over $[234]$ is equivalent to antisymmetrising over $[1234]$, and there is only one completely antisymmetric $k^2 u_1u_2u_3u_4$ scalar.  Without further calculation, one can infer that the kinematic factor is proportional to that scalar,
\be
K^{\text{4F}}_{\text{1-loop}} = \text{const} \cdot \bigl( (u_1 \slashed{k}_3 u_2)(u_3 \slashed{k}_1 u_4) - (u_1 \slashed{k}_2 u_3)(u_2 \slashed{k}_1 u_4) + (u_1 \slashed{k}_2 u_4)(u_2 \slashed{k}_1 u_3) \bigr) \,,
\nn
\ee
which of course agrees with the RNS amplitude (see e.g. \cite{aticksen}, eq. (3.67)).

\subsection{Two bosons, two fermions}
\label{s:oneloop2b2f}

In evaluating \eqref{eq:koneloop} for two bosons and two fermions, the cyclic symmetrisations affect whether the $W$ and ${\cal F}$ superfields contribute bosons or fermions.  Only the label of the $A_\alpha$ superfield stays unaffected, and one has to choose whether it should contribute a boson or a fermion.  Since its fermionic expansion starts with the bosonic polarisation vector, $A_{1,\alpha} \sim (\slashed{\zeta}_1 \theta)_\alpha$, the calculation can be simplified by choosing a labelling where particle 1 is a fermion.  (Of course, the final result must be independent of this choice.)  The assignment of the other three labels is then irrelevant and will be chosen as $f_1f_2b_3b_4$.  Writing out the cyclic permutations, two of the three terms are essentially the same because they are related by interchange of the labels 3 and 4.  The kinematic factor is then
\begin{multline}
K^{\text{2B2F}}_{\text{1-loop}} (f_1f_2b_3b_4) = (1+\pi_{34}) \bigl\langle (\lambda A^{\text{(even)}}_1) (\lambda\gamma^m W^{\text{(even)}}_2) (\lambda\gamma^n W^{\text{(odd)}}_3) {\cal F}^{\text{(even)}}_{4, mn} \bigr\rangle \\
+ \bigl\langle (\lambda A^{\text{(even)}}_1) (\lambda\gamma^m W^{\text{(odd)}}_3) (\lambda\gamma^n W^{\text{(odd)}}_4) {\cal F}^{\text{(odd)}}_{2, mn} \bigr\rangle \,.
\nn
\end{multline}
Unlike in the four-fermion calculation, the result is not fixed by symmetry.  There are five independent $ku_1u_2F_3F_4$ scalars (see appendix \ref{s:spinors}, eq. \eqref{eq:2b2fbasis5}), denoted by $\basisbf{1}\dots\basisbf{5}$, and there are two independent combinations of these scalars with the required $[12](34)$ symmetry.  Expanding the superfields and collecting terms with $\theta^5$, the first line yields a combination of terms $X_{ABCD}$ with $A$, $B$, $D$ odd and $C$ even.  There is only one $\theta^5$ combination coming from the second line, which will be denoted by $X'_{2111}\equiv (-\pi_{24}) X_{2111}$:
\be
K^{\text{2B2F}}_{\text{1-loop}} = (1+\pi_{34}) \left( X_{4010} + X_{2210} + X_{2030} + X_{2012} \right) + X'_{2111} \,,
\nn
\ee
with the correlators
\bea
X_{4010} &= \tfrac{i}{60} k^1_q k^3_b \zeta^3_c k^4_{m} \zeta^4_{n} \tilde{X}_{4010} \,, \;\; &\tilde{X}_{4010} = \bigl\langle (\lambda\gamma^a\theta) (\theta\gamma_a{}^{pq}\theta) (\theta\gamma_p u_1)(\lambda\gamma^{[m} u_2) (\lambda\gamma^{n]}\gamma^{bc}\theta) \bigr\rangle
\nn \\
X_{2210} &= - \tfrac{i}{12} k^2_b k^3_d \zeta^3_e k^4_{m} \zeta^4_{n} \tilde{X}_{2210} \,, \;\; &\tilde{X}_{2210} = \bigl\langle (\lambda\gamma^a\theta)(\theta\gamma_a u_1) (\lambda\gamma^{[m|}\gamma^{bc}\theta) (\theta\gamma_c u_2) (\lambda\gamma^{|n]} \gamma^{de}\theta) \bigr\rangle
\nn \\
X_{2030} &= - \tfrac{i}{36} k^3_b k^3_d \zeta^3_e k^4_{m} \zeta^4_{n} \tilde{X}_{2030} \,, \;\; &\tilde{X}_{2030} = \bigl\langle (\lambda\gamma^a\theta)(\theta\gamma_a u_1) (\lambda\gamma^{[m} u_2) (\lambda\gamma^{n]}\gamma^{bc}\theta) (\theta\gamma_c{}^{de}\theta) \bigr\rangle
\nn \\
X_{2012} &= - \tfrac{i}{12} k^3_b \zeta^3_c k^4_m k^4_d \zeta^4_e \tilde{X}_{2012} \,, \;\; &\tilde{X}_{2012} = \bigl\langle (\lambda\gamma^a\theta)(\theta\gamma_a u_1) (\lambda\gamma^{[m} u_2) (\lambda\gamma^{n]}\gamma^{bc}\theta) (\theta\gamma_n{}^{de}\theta) \bigr\rangle
\nn \\
X'_{2111} &= \tfrac{i}{6} k^3_b \zeta^3_c k^4_d \zeta^4_e k^2_m \tilde{X}'_{2111} \,, \;\; &\tilde{X}'_{2111} = \bigl \langle (\lambda\gamma^a\theta)(\theta\gamma_a u_1) (\lambda\gamma^{[m|}\gamma^{bc}\theta) (\lambda\gamma^{|n]}\gamma^{de}\theta) (\theta\gamma_n u_2) \bigr\rangle
\nn
\eea
(The numerical coefficient in $X'_{2111}$ includes a sign coming from the $\theta$, $\hat{u}$ ordering: there is an odd number of $\theta$s between $u_1$ and $u_2$.)  Evaluating these expressions as outlined in section \ref{s:tracemethod}, the spinor wavefunctions $u_i$ present no complication.  The last part takes the simplest form:  One finds
\be
\bigl\langle (\lambda\gamma^a\theta)(\theta\gamma_a u_1) (\lambda\gamma^{m}\gamma^{bc}\theta) (\lambda\gamma^{n}\gamma^{de}\theta) (\theta\gamma_n u_2) \bigr\rangle = -\tfrac{1}{240} (2\delta^{bc}_{m[d} (u_1 \gamma_{e]} u_2) + \delta^{[b}_m (u_1 \gamma^{c]de} u_2) )
\nn
\ee
and therefore
\be
\tilde{X}'_{2111} = -\tfrac{1}{480} \left( \delta^{[b}_m (u_1 \gamma^{c]}\gamma^{de} u_2) + \delta^{[d}_m (u_1 \gamma^{e]}\gamma^{bc} u_2) \right) \,.
\nn
\ee
The result for $\tilde{X}_{4010}$ is
\bea
\tilde{X}_{4010} &=& \Bigl[ -\tfrac{1}{360} \delta^{bq}_{mn} (u_1 \gamma^c u_2) - \tfrac{1}{90}\delta^{bc}_{mq} (u_1 \gamma^n u_2) + \tfrac{1}{720}\delta^{bc}_{mn} (u_1 \gamma^q u_2) - \tfrac{1}{2520}\delta^m_q (u_1 \gamma^{bcn} u_2) \nn\\
&& - \tfrac{1}{720}\delta^b_q (u_1 \gamma^{cmn} u_2) + \tfrac{1}{1260}\delta^b_m (u_1 \gamma^{cnq} u_2) + \tfrac{1}{3360} (u_1 \gamma^{bcmnq} u_2) \Bigr]_{[bc][mn]} \,.
\nn
\eea
For the evaluation of $\tilde{X}_{2210}$, it is useful to consider the more general correlator
\begin{multline}
\bigl\langle (\lambda\gamma^a\theta)(\theta\gamma_a u_1)(\lambda\gamma^{[m|}\gamma^{bc}\theta)(\lambda\gamma^{|n]}\gamma^{de}\theta) (\theta\gamma_x u_2) \bigr\rangle = \Bigl[ -\tfrac{13}{5040}\delta^d_x\delta^{be}_{mn} (u_1 \gamma^c u_2) + \dots \\
+ \tfrac{11}{201600} \delta^m_x (u_1 \gamma^{bcden} u_2) + \dots - \tfrac{11}{403200} (u_1 \gamma^{bcdemnx} u_2) \bigr]_{[mn][bc][de]} \qquad\text{(27 terms)} \\
+ \tfrac{1}{9676800} \varepsilon_{bcdemni_1i_2i_3i_4} (u_1 \gamma^{i_1i_2i_3i_4x} u_2) - \tfrac{1}{2419200} \varepsilon_{bcdemnxi_1i_2i_3} (u_1 \gamma^{i_1i_2i_3} u_2) \,. \nn
\end{multline}
This time, even using the method of section \ref{s:tracemethod}, there are sufficiently many open indices and long enough traces for epsilon tensors to appear.  Using eqs. \eqref{eq:epsgamma1} and \eqref{eq:epsgamma2}, they can be re-written into $\gamma^{[5,7]}$ terms:
\begin{multline}
\bigl\langle (\lambda\gamma^a\theta)(\theta\gamma_a u_1)(\lambda\gamma^{[m|}\gamma^{bc}\theta)(\lambda\gamma^{|n]}\gamma^{de}\theta) (\theta\gamma_x u_2) \bigr\rangle = \Bigl[ -\tfrac{13}{5040}\delta^d_x\delta^{be}_{mn} (u_1 \gamma^c u_2) + \dots \\
+ \tfrac{1}{16800} \delta^m_x (u_1 \gamma^{bcden} u_2) + \dots - \tfrac{1}{33600} (u_1 \gamma^{bcdemnx} u_2) \Bigr]_{[mn][bc][de]} \qquad\text{(27 terms)} \nn
\end{multline}
A good check on the sign of the epsilon contributions is that $\tilde{X}'_{2111}$ is recovered when contracting with $\eta_{nx}$, involving a cancellation of all $\gamma^{[5]}$ terms.  To obtain $\tilde{X}_{2210}$, one multiplies by $-\eta_{cx}$:
\bea
\tilde{X}_{2210} &=& \Bigl[ \tfrac{1}{720} \delta^{de}_{mn} (u_1 \gamma^b u_2) + \tfrac{29}{2880}\delta^{bd}_{mn} (u_1 \gamma^e u_2) + \tfrac{11}{2880}\delta^{bm}_{de} (u_1 \gamma^n u_2) + \tfrac{1}{20160}\delta^d_m (u_1 \gamma^{ben} u_2) \nn\\
&& + \tfrac{1}{2880}\delta^b_m (u_1 \gamma^{den} u_2) + \tfrac{11}{20160}\delta^b_d (u_1 \gamma^{emn} u_2) + \tfrac{1}{4480} (u_1 \gamma^{bdemn} u_2) \Bigr]_{[de][mn]} \nn
\eea
For the calculation of $X_{2030}$ and $X_{2012}$, one may first evaluate a more general correlator $\langle (\lambda\gamma^a\theta)(\theta\gamma_a u_1)(\lambda\gamma^{[m} u_2)(\lambda\gamma^{n]}\gamma^{bc}\theta)(\theta\gamma^x\gamma^{de}\theta) \rangle$ and then contract with $\eta_{cx}$ and $\eta_{nx}$, respectively.  The results are
\bea
\tilde{X}_{2030} &=& \Bigl[ -\tfrac{1}{720} \delta^{de}_{mn} (u_1 \gamma^b u_2) + \tfrac{1}{288}\delta^{bd}_{mn} (u_1 \gamma^e u_2) - \tfrac{1}{1440}\delta^{bm}_{de} (u_1 \gamma^n u_2) - \tfrac{17}{10080}\delta^d_m (u_1 \gamma^{ben} u_2) \nn\\
&& - \tfrac{23}{10080}\delta^b_m (u_1 \gamma^{den} u_2) - \tfrac{1}{1440}\delta^b_d (u_1 \gamma^{emn} u_2) + \tfrac{1}{6720} (u_1 \gamma^{bdemn} u_2) \Bigr]_{[mn][de]} \,,
\nn\\
\tilde{X}_{2012} &=& \Bigl[ \tfrac{1}{288} \delta^{de}_{bm} (u_1\gamma^c u_2) + \tfrac{1}{288}\delta^{bc}_{dm} (u_1 \gamma^e u_2) - \tfrac{1}{1440}\delta^{bc}_{de} (u_1 \gamma^m u_2) + \tfrac{1}{2016}\delta^d_m (u_1 \gamma^{bce} u_2) \nn\\
&& - \tfrac{11}{10080}\delta^b_m (u_1 \gamma^{cde} u_2) + \tfrac{17}{10080}\delta^b_d (u_1 \gamma^{cem} u_2) - \tfrac{1}{3360} (u_1 \gamma^{bcdem} u_2) \Bigr]_{[bc][de]} \,.
\nn
\eea
After multiplication with the momenta and polarisations,  all individual contributions are gauge invariant and can be expanded in the basis $\basisbf{1}\dots\basisbf{5}$ listed in \eqref{eq:2b2fbasis5}:
\bea
(1 + \pi_{34}) X_{4010} &=& \tfrac{i}{483840}(-6, -16, -40, 6, 0)_{\basisbf{1}\dots\basisbf{5}} \nn\\
(1 + \pi_{34}) X_{2210} &=& \tfrac{i}{483840}(-18, -104, -176, 18, 0)_{\basisbf{1}\dots\basisbf{5}} \nn\\
(1 + \pi_{34}) X_{2030} &=& \tfrac{i}{483840}(-21, 42, -42, 21, 0)_{\basisbf{1}\dots\basisbf{5}} \nn\\
(1 + \pi_{34}) X_{2012} &=& \tfrac{i}{483840}(-39, 78, -78, 39, 0)_{\basisbf{1}\dots\basisbf{5}} \nn\\
X'_{2111} &=& -\tfrac{i}{11520}(1,0,4,-1,0)_{\basisbf{1}\dots\basisbf{5}} \nn
\eea
The sum can be written as
\bea
K^{\text{2B2F}}_{\text{1-loop}} = X'_{2111} &=& -\tfrac{i}{3840} (1,0,4,-1,0)_{\basisbf{1}\dots\basisbf{5}} \nn\\
&=& -\tfrac{i}{1920} \Bigl( \kk{1}{3} (u_2 \slashed{\zeta}_3 (\slashed{k}_2+\slashed{k}_3) \slashed{\zeta}_4 u_1) + \kk{2}{3} (u_2 \slashed{\zeta}_4 (\slashed{k}_2+\slashed{k}_4) \slashed{\zeta}_3 u_1) \Bigr)
\eea
and again agrees with the amplitude computed in the RNS result, see \cite{aticksen} eq. (3.37).

\section{Two-loop amplitudes}
\label{s:twoloop}

The pure spinor formalism was used in \cite{btwoloop,bmultiloop} to compute the two-loop type-IIB amplitude involving four massless states,
\bea
{\cal A} &=& \int d^2 \Omega_{11} d^2 \Omega_{12} d^2 \Omega_{22} \prod_{i=1}^4 \int d^2 z_i \frac{\exp \left( -\sum_{i,j} k_i\cdot k_j \, G(z_i, z_j) \right)}{(\det \Imag \Omega)^5} K_{\text{2-loop}} (k_i, z_i) \,,
\nn
\eea
where $\Omega$ is the genus-two period matrix, and the integration over fermionic zero modes is encapsulated in
\bea
\label{eq:ktwoloop}
K_{\text{2-loop}} &=& \Delta_{12}\Delta_{34} \bigl\langle (\lambda\gamma^{mnpqr}\lambda)(\lambda\gamma^s W_1) {\cal F}_{2,mn}{\cal F}_{3,pq}{\cal F}_{4,rs} \bigr\rangle + \bigl(\text{perm(1234)}\bigr)
\\
\label{eq:kijtwoloop}
&\equiv& \Delta_{12}\Delta_{34} K_{12} + \Delta_{13}\Delta_{24} K_{13} + \Delta_{14}\Delta_{23} K_{14} \,.
\eea
The kinematic factors $K_{12}$, $K_{13}$, $K_{14}$ are accompanied by the basic antisymmetric biholomorphic 1-form $\Delta$, which is related to a canonical basis $\omega_1, \omega_2$ of holomorphic differentials via $\Delta_{ij} = \Delta(z_i,z_j) = \omega_1(z_i)\omega_2(z_j) - \omega_2(z_i)\omega_1(z_j)$.  The superfields $W_i^\alpha$ and ${\cal F}_{i,mn}$ are the spinor and vector field strengths of the $i$-th external state, as in section \ref{s:oneloop}.  One encounters superspace integrals of the form
\be
\label{eq:yabcd}
Y(abcd) = \bigl\langle (\lambda\gamma^{mnpqr}\lambda)(\lambda\gamma^s W_a) {\cal F}_{b,mn} {\cal F}_{c,pq} {\cal F}_{d,rs} \bigr\rangle \,.
\ee
The symmetries of the $\lambda^3$ combination \cite{btwoloop} in this correlator include the obvious symmetry under $mn\leftrightarrow pq$, and also $(\lambda\gamma^{[mnpqr}\lambda)(\lambda\gamma^{s]})_{\alpha} = 0$ (this holds for pure spinors $\lambda$ and can be seen by dualising, and holds for unconstrained spinors $\lambda$ as part of a $\lambda^3\theta^5$ scalar, as seen from the representation content \eqref{eq:la3th5reps}), and allow one to shuffle the ${\cal F}$ factors:
\be
Y(abcd)=Y(acbd) \,, \qquad Y(abcd) + Y(acdb) + Y(adbc) = 0 \,.
\label{eq:yshuffle}
\ee

\subsection{Review: four bosons}
\label{s:twoloop4b}

The case of four Neveu-Schwarz states was considered in \cite{bmtwoloop} and will be briefly reviewed here.  As all three kinematic factors $K_{12}$, $K_{13}$ and $K_{14}$ are equivalent, it is sufficient to consider $K_{12}$ in detail.  With all external states being identical, the symmetrisations of \eqref{eq:ktwoloop} can be carried out at the end of the calculation:
\bea
K^{\text{4B}}_{12} &=& {} 4 \left.\bigl\langle W_{[1} {\cal F}_{2]} {\cal F}_{[3} {\cal F}_{4]} \bigr\rangle\right|_{\text{4B}} + 4 \left.\bigl\langle W_{[3} {\cal F}_{4]} {\cal F}_{[1} {\cal F}_{2]} \bigr\rangle\right|_{\text{4B}} \nn\\
&=& (1-\pi_{12})(1-\pi_{34})(1+\pi_{13}\pi_{24}) \left.\bigl\langle W_{1} {\cal F}_{2} {\cal F}_{3} {\cal F}_{4} \bigr\rangle\right|_{\text{4B}}
\nn
\eea
Expanding the superfields and adopting the notation
\be
\nn
Y_{ABCD}(abcd) = \left\langle (\lambda\gamma^{mnpqr}\lambda)(\lambda\gamma^s W_a^{(A)}) {\cal F}_{b,mn}^{(B)} {\cal F}_{c,pq}^{(C)} {\cal F}_{d,rs}^{(D)} \right\rangle \,,
\ee
the Neveu-Schwarz states come from terms of the form $Y_{ABCD} \equiv Y_{ABCD}(1234)$ with $A$~odd and $B$, $C$, $D$ even.  Using the shuffling identities \eqref{eq:yshuffle} to simplify, one obtains
\begin{multline}
\left.\bigl\langle W_{1} {\cal F}_{2} {\cal F}_{3} {\cal F}_{4} \bigr\rangle \right|_{\text{4B}} \\
\begin{split}
&= Y_{5000}+Y_{1400}+Y_{1040}+Y_{1004}+Y_{3200}+Y_{3020}+Y_{3002}+Y_{1220}+Y_{1202}+Y_{1022} \nn\\
&= (1+\pi_{23})(1-\pi_{24}) \left( \tfrac{1}{3} Y_{5000} + Y_{1400} + Y_{3200} + Y_{1022} \right) \,,
\end{split}
\nn
\end{multline}
and therefore $K^{\text{4B}}_{12}$ can be written as the image of a symmetrisation operator ${\cal S_{\text{4B}}}$:
\bea
K^{\text{4B}}_{12} &=& {\cal S}_{\text{4B}} \left( \tfrac{1}{3} Y_{5000} + Y_{1400} + Y_{3200} + Y_{1022} \right) \nn \\
{\cal S_{\text{4B}}} &=& (1-\pi_{12})(1-\pi_{34})(1+\pi_{13}\pi_{24})(1+\pi_{23})(1-\pi_{24})
\nn
\eea
It is worth noting at this point that, on the sixteen-dimensional space of Lorentz scalars built from the four field strengths $F_i$ and two momenta, the symmetriser ${\cal S}_{\text{4B}}$ has rank four.  The correlators were computed in \cite{bmtwoloop}, using the method outlined in section \ref{s:tensorial}.  Two are zero, $Y_{5000}=Y_{1400}=0$, and the remaining ones are
\bea
Y_{3200} &=& \tfrac{1}{192} k^1_a F^1_{cd} k^2_m F^2_{ef} F^3_{pq} F^4_{rs} \; \bigl\langle (\lambda\gamma^{mnpqr}\lambda) (\lambda\gamma^s \gamma^{ab} \theta) (\theta\gamma_b{}^{cd}\theta)(\theta\gamma_n{}^{ef}\theta) \bigr\rangle \,,
\nn \\
Y_{1022} &=& \tfrac{1}{64} F^1_{ab} F^2_{mn} k^3_p F^3_{cd} k^4_r F^4_{ef} \; \bigl\langle (\lambda\gamma^{mnpq[r}\lambda) (\lambda\gamma^{s]} \gamma^{ab} \theta) (\theta\gamma_q{}^{cd}\theta)(\theta\gamma_s{}^{ef}\theta) \bigr\rangle \,.
\nn
\eea
In reducing those two contributions to a set of independent scalars, one finds that they both are not just sums of $(k\cdot k) F^4$ terms but also contain terms of the form $k\cdot F$ terms.  The latter are projected out by the symmetriser ${\cal S}_{\text{4B}}$, and the result is
\bea
K^{\text{4B}}_{12} = {\cal S}_{\text{4B}} (Y_{3200}+Y_{1022}) &=& \tfrac{1}{120} (\kk{1}{3} - \kk{2}{3}) \left(4 \Tr(F_{(1}F_2F_3F_{4)})- \Tr(F_{(1}F_2) \Tr(F_3F_{4)}) \right) \,, \nn \\
 &=& \tfrac{1}{720} (\kk{1}{3} - \kk{2}{3}) t_8 F^4 \,.
\nn
\eea
By trivial index exchange, one obtains $K_{13}$ and $K_{14}$, and the total is
\be
K^{\text{4B}}_{\text{2-loop}} = \tfrac{1}{720} \bigl( (\kk{1}{3} - \kk{2}{3}) \Delta_{12}\Delta_{34} + (\kk{1}{2}-\kk{2}{3}) \Delta_{13}\Delta_{24} + (\kk{1}{2}-\kk{1}{3}) \Delta_{14}\Delta_{23} \bigr) t_8 F^4 \,, \label{eq:2loop4b}
\ee
a product of the completely symmetric one-loop kinematic factor $t_8 F^4$ and a completely symmetric combination of the momenta and the $\Delta_{ij}$.

\subsection{Four fermions}
\label{s:twoloop4f}

The calculation involving four Ramond states is very similar to the bosonic one.  Focussing on the $K_{12}$ part, the symmetrisations in \eqref{eq:ktwoloop} can again be rewritten as action of symmetrisation operators on the correlator of superfields with one particular labelling:
\bea
K^{\text{4F}}_{12} (\hat{u}_i) &=& (1-\pi_{12})(1-\pi_{34})(1+\pi_{13}\pi_{24}) \left.\bigl\langle W_{1} {\cal F}_{2} {\cal F}_{3} {\cal F}_{4} \bigr\rangle\right|_{\hat{u}_1\hat{u}_2\hat{u}_3\hat{u}_4} \nn \\
&=& 4 (1 - \pi_{12}) \left.\bigl\langle W_{1} {\cal F}_{2} {\cal F}_{3} {\cal F}_{4} \bigr\rangle\right|_{\hat{u}_1\hat{u}_2\hat{u}_3\hat{u}_4} \nn
\eea
The last step follows from the fact that all scalars of the form $k^4u^4$ (see appendix \ref{s:basis4f}), and therefore all $k^4\hat{u}^4$ scalars, are invariant under $\pi_{13}\pi_{24}$ and have $\pi_{12}=\pi_{34}$.  This time, on expanding the superfields, one collects the terms $Y_{ABCD}$ with $A$ even and $B$, $C$, $D$ odd.  After using \eqref{eq:yshuffle} to simplify,
\bea
\left.\bigl\langle W_{1} {\cal F}_{2} {\cal F}_{3} {\cal F}_{4} \bigr\rangle \right|_{\hat{u}_1\hat{u}_2\hat{u}_3\hat{u}_4} &=& Y_{2111} + Y_{0311} + Y_{0131} + Y_{0113} \nn\\
&=& (1+\pi_{23})(1-\pi_{24}) \left( \tfrac{1}{3} Y_{2111} + Y_{0311} \right) \,,
\nn
\eea
and after translating to commuting wavefunctions $u_i$, which multiplies every permutation operator with its signature, one obtains
\be
K^{\text{4F}}_{12} (u_i) = {\cal S}_{\text{4F}} \left( \tfrac{1}{3} Y_{2111} (u_i) + Y_{0311} (u_i) \right) \,, \qquad {\cal S}_{\text{4F}} = 4 (1 + \pi_{12})(1-\pi_{23})(1+\pi_{24}) \,.
\nn
\ee
This symmetriser has rank three, and the result is again not determined by symmetry.  Two correlators have to be computed:
\bea
Y_{2111}(u_i) &=& (-2) k^1_a k^2_m k^3_p k^4_r \; \bigl\langle (\lambda \gamma^{mnpq[r} \lambda) (\lambda \gamma^{s]}\gamma^{ab} \theta) (\theta\gamma_b u_1) (\theta\gamma_n u_2) (\theta\gamma_q u_3) (\theta\gamma_s u_4) \bigr\rangle
\nn\\
Y_{0311}(u_i) &=& (-\tfrac{2}{3}) k^2_a k^2_m k^3_p k^4_r \; \bigl\langle (\lambda\gamma^{mnpq[r} \lambda) (\lambda\gamma^{s]} u_1) (\theta\gamma_n{}^{ab} \theta) (\theta\gamma_b u_2) (\theta\gamma_q u_3) (\theta\gamma_s u_4) \bigr\rangle
\nn
\eea
With four fermions present, the method of section \ref{s:tracemethod} is preferred as it does not involve re-arranging the fermions using Fierz identities.  The first correlator was covered as an example in that section, and the second one can be evaluated in the same fashion.  Expressed in the basis listed in \eqref{eq:4fbasis10}, the results are
\bea
Y_{2111}(u_i) &=& \tfrac{1}{5040}(-19,-21,21,19,-17,-17,0,0,0,0)_{\basisf{1}\dots\basisf{10}} \,, \nn \\
Y_{0311}(u_i) &=& \tfrac{1}{15120}(-14,-16,0,2,-8,-8,0,-5,-5,0)_{\basisf{1}\dots\basisf{10}} \,. \nn
\eea
After acting with the symmetriser ${\cal S}_{\text{4F}}$, one obtains the same $u^4$ scalar encountered in the one-loop amplitude,
\bea
K^{\text{4F}}_{12} (u_i) &=& {\cal S}_{\text{4F}}(\tfrac{1}{3} Y_{2111}(u_i) + Y_{0311}(u_i)) = \tfrac{1}{45} (-1,-2,1,2,-1,-2,0,0,0,0)_{\basisf{1}\dots\basisf{10}} \nn\\
&=& \tfrac{1}{45} (\kk{2}{3} - \kk{1}{3}) \, \bigl( (u_1 \slashed{k}_3 u_2)(u_3 \slashed{k}_1 u_4) - (u_1 \slashed{k}_2 u_3)(u_2 \slashed{k}_1 u_4) + (u_1 \slashed{k}_2 u_4)(u_2 \slashed{k}_1 u_3) \bigr) \,.
\nn
\eea
The $K_{13}$ and $K_{14}$ parts again follow by index exchange, and the total result
\begin{multline}
K^{\text{4F}}_{\text{2-loop}} (u_i) = \tfrac{1}{45} \bigl( (\kk{2}{3}-\kk{1}{3}) \Delta_{12}\Delta_{34} + (\kk{2}{3}-\kk{1}{2}) \Delta_{13}\Delta_{24} + (\kk{1}{3}-\kk{1}{2}) \Delta_{14}\Delta_{23} \bigr) \\
\times \bigl( (u_1 \slashed{k}_3 u_2)(u_3 \slashed{k}_1 u_4) - (u_1 \slashed{k}_2 u_3)(u_2 \slashed{k}_1 u_4) + (u_1 \slashed{k}_2 u_4)(u_2 \slashed{k}_1 u_3) \bigr) \label{eq:2loop4f}
\end{multline}
is again a simple product of the one-loop kinematic factor and a combination of the $\Delta_{ij}$ and momenta.

\subsection{Two bosons, two fermions}
\label{s:twoloop2b2f}

As in the one-loop calculation of section \ref{s:oneloop2b2f}, in the mixed case one has to pay some attention to the permutations in \eqref{eq:ktwoloop} since they affect which superfields contribute fermionic fields.  The complete symmetrisation makes it irrelevant which labels are assigned to the two fermions, and the convention $f_1f_2b_3b_4$ will be used here.  The kinematic factor $K^{\text{2B2F}}_{12}$ is then distinguished from the other two, $K^{\text{2B2F}}_{13}$ and $K^{\text{2B2F}}_{14}$.  Carrying out the symmetrisations in \eqref{eq:ktwoloop} and using the identities \eqref{eq:yshuffle}, one finds
\bea
K_{12} (\hat{u}_1,\hat{u}_2,\zeta_3,\zeta_4) &=& ({\mathbf 1} - \pi_{12}) ({\mathbf 1} - \pi_{34}) \tilde{K} \,, \nn \\
K_{13} (\hat{u}_1,\hat{u}_2,\zeta_3,\zeta_4) &=& (2 \cdot {\mathbf 1} + \pi_{12} + \pi_{34} + 2 \pi_{12} \pi_{34} ) \tilde{K} \,, \nn \\
K_{14} (\hat{u}_1,\hat{u}_2,\zeta_3,\zeta_4) &=& ({\mathbf 1} + 2 \pi_{12} + 2 \pi_{34} + \pi_{12} \pi_{34} ) \tilde{K} \,, \nn
\eea
where, schematically,
\be
\tilde{K} = \bigl\langle (\lambda^3 W^{\text{(even)}}_1) {\cal F}^{\text{(odd)}}_2 {\cal F}^{\text{(even)}}_3 {\cal F}^{\text{(even)}}_4 \bigr\rangle + \bigl\langle (\lambda^3 W^{\text{(odd)}}_3) {\cal F}^{\text{(even)}}_4 {\cal F}^{\text{(odd)}}_1 {\cal F}^{\text{(odd)}}_2 \bigr\rangle \,.
\label{eq:ktilde2b2f}
\ee
In translating to commuting variables $u_1$ and $u_2$, the permutation operator $\pi_{12}$ changes sign, and therefore\footnote{This sign change is crucial to avoid the erroneous conclusion that the two-boson, two-fermion kinematic factor cannot be of the same product form as in the four-boson or four-fermion cases, which would be in contradiction to the supersymmetric identities derived in \cite{mafraid}.}
\bea
K_{12} (u_1,u_2,\zeta_3,\zeta_4) &=& ({\mathbf 1} + \pi_{12}) ({\mathbf 1} - \pi_{34}) \tilde{K} \,, \nn \\
K_{13} (u_1,u_2,\zeta_3,\zeta_4) &=& (2 \cdot {\mathbf 1} - \pi_{12} + \pi_{34} - 2 \pi_{12} \pi_{34} ) \tilde{K} \,, \nn \\
K_{14} (u_1,u_2,\zeta_3,\zeta_4) &=& ({\mathbf 1} - 2 \pi_{12} + 2 \pi_{34} - \pi_{12} \pi_{34} ) \tilde{K} \,. \nn
\eea
Expanding the superfields, the contributions to $\tilde{K}$ are:
\bea
Y_{4100} &=& -\tfrac{i}{48} k^1_a k^1_d k^2_m F^3_{pq} F^4_{rs} \; \bigl\langle (\lambda \gamma^{mnpqr} \lambda) (\lambda \gamma^{s} \gamma^{ab} \theta) (\theta\gamma_b \gamma^{cd} \theta) (\theta\gamma_c u_1) (\theta\gamma_n u_2) \bigr\rangle
\nn\\
Y_{0500} &=& \tfrac{i}{240} k^2_m k^2_a k^2_c F^3_{pq} F^4_{rs} \; \bigl\langle (\lambda \gamma^{mnpqr} \lambda) (\lambda \gamma^{s} u_1) (\theta\gamma_n{}^{ab} \theta) (\theta\gamma_b{}^{cd} \theta) (\theta\gamma_d u_2) \bigr\rangle
\nn\\
Y_{0140} &=& \tfrac{i}{48} k^2_m k^3_p k^3_a F^3_{cd} F^4_{rs} \; \bigl\langle (\lambda \gamma^{mnpqr} \lambda) (\lambda \gamma^{s} u_1) (\theta\gamma_n u_2) (\theta\gamma_q{}^{ab} \theta) (\theta\gamma_b{}^{cd}\theta) \bigr\rangle
\nn\\
Y_{0104} &=& \tfrac{i}{48} k^2_m F^3_{pq} k^4_a F^4_{cd} k^4_{[r|} \; \bigl\langle (\lambda \gamma^{mnpqr} \lambda) (\lambda \gamma^{s} u_1) (\theta\gamma_n u_2) (\theta\gamma_{|s]}{}^{ab} \theta) (\theta\gamma_b{}^{cd}\theta) \bigr\rangle
\nn\\
Y_{2300} &=& \tfrac{i}{24} k^1_a k^2_m k^2_c F^3_{pq} F^4_{rs} \; \bigl\langle (\lambda \gamma^{mnpqr} \lambda) (\lambda \gamma^{s}\gamma^{ab} \theta) (\theta\gamma_b u_1) (\theta\gamma_n{}^{cd}\theta) (\theta\gamma_e u_2) \bigr\rangle
\nn\\
Y_{2120} &=& \tfrac{i}{8} k^1_a k^2_m k^3_p F^3_{cd} F^4_{rs} \; \bigl\langle (\lambda \gamma^{mnpqr} \lambda) (\lambda \gamma^{s}\gamma^{ab} \theta) (\theta\gamma_b u_1) (\theta\gamma_n u_2) (\theta\gamma_q{}^{cd}\theta) \bigr\rangle
\nn\\
Y_{2102} &=& \tfrac{i}{8} k^1_a k^2_m F^3_{pq} F^4_{cd} k^4_{[r|} \; \bigl\langle (\lambda \gamma^{mnpqr} \lambda) (\lambda \gamma^{s}\gamma^{ab} \theta) (\theta\gamma_b u_1) (\theta\gamma_n u_2) (\theta\gamma_{|s]}{}^{cd}\theta) \bigr\rangle
\nn\\
Y_{0320} &=& \tfrac{i}{24} k^2_m k^2_a k^3_p F^3_{cd} F^4_{rs} \; \bigl\langle (\lambda \gamma^{mnpqr} \lambda) (\lambda \gamma^{s} u_1) (\theta\gamma_n{}^{ab}\theta) (\theta\gamma_b u_2) (\theta\gamma_q{}^{cd}\theta) \bigr\rangle
\nn\\
Y_{0302} &=& \tfrac{i}{24} k^2_m k^2_a F^3_{pq} F^4_{cd} k^4_{[r|} \; \bigl\langle (\lambda \gamma^{mnpqr} \lambda) (\lambda \gamma^{s} u_1) (\theta\gamma_n{}^{ab}\theta) (\theta\gamma_b u_2) (\theta\gamma_{|s]}{}^{cd}\theta) \bigr\rangle
\nn\\
Y_{0122} &=& \tfrac{i}{4} k^2_m k^3_p F^3_{ab} F^4_{cd} k^4_{[r|} \; \bigl\langle (\lambda \gamma^{mnpqr} \lambda) (\lambda \gamma^{s} u_1) (\theta\gamma_n u_2) (\theta\gamma_q{}^{ab}\theta) (\theta\gamma_{s]}{}^{cd}\theta) \bigr\rangle
\nn\\
Y_{3011} &=& \tfrac{i}{12} k^3_a F^3_{cd} F^4_{mn} k^1_p k^2_{[r|} \; \bigl\langle (\lambda \gamma^{mnpqr} \lambda) (\lambda \gamma^{s}\gamma^{ab}\theta) (\theta\gamma_b{}^{cd}\theta)(\theta\gamma_c u_1) (\theta\gamma_n u_2) \bigr\rangle
\nn\\
Y_{1211} &=& \tfrac{i}{2} F^3_{ab} k^4_m F^4_{cd} k^1_p k^2_{[r|} \; \bigl\langle (\lambda \gamma^{mnpqr} \lambda) (\lambda \gamma^{s}\gamma^{ab}\theta) (\theta\gamma_n{}^{cd}\theta) (\theta\gamma_q u_1) (\theta\gamma_{|s]} u_2) \bigr\rangle
\nn\\
Y_{1031} &=& \tfrac{i}{12} F^3_{ab} F^4_{mn} k^1_p k^1_c k^2_{[r|} \; \bigl\langle (\lambda \gamma^{mnpqr} \lambda) (\lambda \gamma^{s}\gamma^{ab}\theta) (\theta\gamma_q{}^{cd}\theta) (\theta\gamma_d u_1) (\theta\gamma_{|s]} u_2) \bigr\rangle
\nn\\
Y_{1013} &=& \tfrac{i}{12} F^3_{ab} F^4_{mn} k^1_p k^2_c k^2_{[r|} \; \bigl\langle (\lambda \gamma^{mnpqr} \lambda) (\lambda \gamma^{s}\gamma^{ab}\theta) (\theta\gamma_q u_1) (\theta\gamma_{|s]}{}^{cd}\theta) (\theta\gamma_d u_2) \bigr\rangle
\nn
\eea
These correlators can be evaluated exactly as described in section \ref{s:oneloop2b2f}.
One finds that $Y_{0500}=Y_{0140}=Y_{0104}=0$, and the sum of the remaining terms reduces to
\bea
\tilde{K} &=& Y_{4100} + Y_{2300} + Y_{2120} + Y_{2102} + Y_{0320} + Y_{0302} + Y_{0122} + Y_{3011} + Y_{1211} + Y_{1031} + Y_{1013} \nn \\
 &=& \tfrac{i}{360}(s_{12}+s_{13}) \times (1,0,4,-1,0)_{C_1\dots C_5} \,. \nn
\eea
After applying the symmetrisation operators,
\bea
({\mathbf 1} + \pi_{12}) ({\mathbf 1} - \pi_{34}) \tilde{K}
&=& \; \tfrac{i}{180} (s_{12}+2s_{13}) \times (1,0,4,-1,0)_{C_1\dots C_5} \,, \nn \\
(2 \cdot {\mathbf 1} - \pi_{12} + \pi_{34} - 2 \pi_{12} \pi_{34} ) \tilde{K}
&=& \; \tfrac{i}{180} (2s_{12}+s_{13}) \times (1,0,4,-1,0)_{C_1\dots C_5} \,, \nn \\
({\mathbf 1} - 2 \pi_{12} + 2 \pi_{34} - \pi_{12} \pi_{34} ) \tilde{K}
&=& \; \tfrac{i}{180} (s_{12}-s_{13}) \times (1,0,4,-1,0)_{C_1\dots C_5} \,, \nn
\eea
the total kinematic factor is seen to be
\begin{multline}
K_{\text{2-loop}} (u_1,u_2,\zeta_3,\zeta_4) = -\tfrac{i}{180} \bigl( (s_{23}-s_{13}) \Delta_{12}\Delta_{34} + (s_{23}-s_{12}) \Delta_{12}\Delta_{34} + (s_{13}-s_{12}) \Delta_{12}\Delta_{34} \bigr) \\
  \times (1,0,4,-1,0)_{C_1\dots C_5} \label{eq:2loop2b2f}
\end{multline}
and displays the same simple product form as in the four-boson and four-fermion case.

\section{Discussion}

In this paper, different methods were discussed to efficiently evaluate the superspace integrals appearing in multiloop amplitudes derived in the pure spinor formalism.  Extending previous calculations \cite{bmtwoloop,moneloop} restricted to Neveu-Schwarz states, it was then shown how the treatment of Ramond states poses no additional difficulties.

While the bosonic calculations of \cite{bmtwoloop,moneloop} have, in conjunction with supersymmetry, already established the equivalence of the massless four-point amplitudes derived in the pure spinor and RNS formalisms, it would be interesting to make contact between the results of sections \ref{s:twoloop4f} / \ref{s:twoloop2b2f} and two-loop amplitudes involving Ramond states as computed in the RNS formalism (see for example \cite{zhu2b2f}).

The assistance of a computer algebra system seems indispensible in explicitly evaluating pure spinor superspace integrals.  To avoid excessive use of custom-made algorithms, it would be desirable to implement these calculations in a wider computational framework particular adapted to field theory calculations \cite{cadabra}.

The methods outlined in this paper should be easily applicable to future higher-loop amplitude expressions derived from the pure spinor formalism, and, it is hoped, to other superspace integrals.

\section*{Acknowledgements}

The author would like to thank Louise Dolan for discussions, and Carlos Mafra for valuable correspondence.  This work is supported by the U.S.~Department of Energy, grant no.~DE-FG01-06ER06-01, Task~A.

\newpage

\appendix

\section{Reduction to kinematic bases}
\label{s:spinors}

In calculating scattering amplitudes one encounters kinematic factors which are Lorentz invariant polynomials in the momenta, polarisations and/or spinor wavefunctions of the scattered particles.  It can be a non-trivial task to simplify such expressions, taking into account the on-shell identities $\sum_i k_i = 0$, $k_i^2 = 0$, $k_i\cdot \zeta_i = 0$, $\slashed{k}_i u_i = 0$, and, in the case of fermions, re-arrangements stemming from Fierz identities.

More generally, one would like to know how many independent combinations of some given fields (subject to on-shell identitites) there are, and how to reduce an arbitrary expression with respect to some chosen basis.  This appendix outlines methods to address these problems, with an emphasis on algorithms which can easily be transferred to a computer algebra system.  These methods are not limited to dealing with pure spinor calculations but the scope will be restricted to amplitudes of four massless vector or spinor particles in ten dimensions.

\subsection{Four bosons}

It is not difficult to reduce polynomials in the momenta and polarisations to a canonical form.  The momentum conservation constraint $\sum_i k_i=0$ is solved by eliminating one momentum (for example $k_4$), all $k_i^2$ are set to zero, and one of the two remaining quadratic combinations of momenta is eliminated (for example $\kk{2}{3} \to - \kk{1}{2} - \kk{1}{3}$, where $\kk{i}{j}\equiv k_i\cdot k_j$).  Then all products $k_i\cdot \zeta_i$ are set to zero, and one extra $k\cdot\zeta$ product is replaced (when eliminating $k_4$, the replacement is $k_3\cdot \zeta_4\to (-k_1-k_2)\cdot\zeta_4$).  The remaining monomials are then independent.  (This is at least the case with the low powers of momenta encountered in the calculations of sections \ref{s:oneloop} and \ref{s:twoloop}, where there are enough spatial directions for all momenta/polarisations to be linearly independent.)

The implementation of these reduction rules on a computer is straightforward.  The easiest way to obtain scalars which are also invariant under the gauge symmetry $k_i\to \zeta_i$ is to start with expressions constructed from the field strengths $F_i^{ab} = 2 \partial^{[a}\zeta_i^{b]}$.  For the one-loop calculations of section \ref{s:oneloop4b}, the relevant basis consists of gauge invariant scalars containing only the four field strengths $F_1\dots F_4$.  One finds six independent combinations,
\begin{gather}
\begin{split}
&\Tr(F_1F_2F_3F_4) \\ &\Tr(F_1F_2F_4F_3) \\ &\Tr(F_1F_3F_2F_4)
\end{split}
\qquad\qquad
\begin{split}
&\Tr(F_1F_2)\Tr(F_3F_4) \\ &\Tr(F_1F_3)\Tr(F_2F_4) \\ &\Tr(F_1F_4)\Tr(F_2F_3)
\end{split}
\nn
\end{gather}
In the two-loop calculations of section \ref{s:twoloop4b}, all monomials have two more momenta.  There are sixteen independent gauge invariant scalars of the form $kkF_1F_2F_3F_4$, and twelve of them may be constructed from the previous basis by multiplication with $\kk{1}{2}$ and $\kk{1}{3}$: $\basisb{1}=\kk{1}{2} \Tr(F_1F_2F_3F_4)$, $\basisb{2}=\kk{1}{3} \Tr(F_1F_2F_3F_4)$, etc.  One choice for the additional four is
\bea
\basisb{13} = k_3\cdot F_1\cdot F_2\cdot k_3 \Tr(F_3F_4) \qquad && \basisb{15} =k_3\cdot F_1\cdot F_4\cdot k_2 \Tr(F_2F_3) \nn\\
\basisb{14} = k_4\cdot F_1\cdot F_3\cdot k_2 \Tr(F_2F_4) \qquad && \basisb{16} =k_4\cdot F_2\cdot F_3\cdot k_4 \Tr(F_1F_4) \nn\,.
\eea
As an example application of the computer algorithms, one may check that the symmetrisation operator of section \ref{s:twoloop4b},
\bea
{\cal S}_{\text{4B}} &=& (1-\pi_{12})(1-\pi_{34})(1+\pi_{13}\pi_{24})(1+\pi_{23})(1-\pi_{24}) \,,
\nn
\eea
acts as
\bea
{\cal S}_{\text{4B}} \basisb{1} &=& 8 \basisb{1} + 4 \basisb{2} - 4 \basisb{3} + 4 \basisb{4} + 8 \basisb{5} + 16 \basisb{6} \nn \\
&\dots& \nn \\
{\cal S}_{\text{4B}} \basisb{16} &=& -6 \basisb{1} + 6 \basisb{3} - 6 \basisb{5} - 12 \basisb{6} + \tfrac{3}{2} \basisb{7} + 3 \basisb{8} + \tfrac{3}{2} \basisb{9} + 3 \basisb{10} + \tfrac{3}{2} \basisb{11} + 3 \basisb{12} \nn
\eea
and has rank four.

\subsection{Four fermions}
\label{s:basis4f}

In dealing with the spinor wavefunctions $u_i$ one has to face two issues: Fierz identities, and the Dirac equation.  Fierz identities not only allow one to change the order of the spinors but also give rise to relations between different expressions in one spinor order.  The Dirac equation often simplifies terms with momenta contracted into $(u_i \gamma^{[n]} u_j)$ bilinears.

In this section it is shown how to construct bases for terms of the form $(k^2\;\text{or}\;k^4) \times u_1u_2u_3u_4$.  A significant simplification comes from noting that the Dirac equation allows one to rewrite $(u_i \gamma^{[n]} u_j)$ bilinears into terms with lower $n$ if more than one momentum is contracted into the $\gamma^{[n]}$.  A good first step is therefore to disregard the momenta temporarily and find all independent scalars and two-index tensors built from $u_1, \dots, u_4$.  From the SO(10) representation content,
\be
(S^+)^{\otimes 4} = 2 \cdot {\mathbf 1} + 6 \cdot \tableau{1 1} + 3 \cdot \widetilde{\tableau{2}} + (\text{tensors with rank $>2$}) \,,\nn
\ee 
one expects two scalars and nine 2-tensors.  The scalars are easily found by considering, as in \cite{gsw},
\bea
T_1(1234) &=& (u_1 \gamma^a u_2)(u_3 \gamma_a  u_4) \,, \nn\\
T_3(1234) &=& (u_1 \gamma^{abc} u_2)(u_3 \gamma_{abc}  u_4) \,.\nn
\eea
and similarly for the other two inequivalent orders of the four spinors.  (Note there is no $T_5$ because of self-duality of the $\gamma^{[5]}$.)  From Fierz transformations, one learns that all $T_3$ terms can be reduced to $T_1$ by $T_3(1234) = -12 T_1(1234) - 24 T_1(1324)$ and permutations, and the identity $(\gamma_a)_{(\alpha\beta}(\gamma^a)_{\gamma)\delta}=0$ implies that $T_1(1234)+T_1(1324)+T_1(1423)=0$, leaving for example $T_1(1234)$ and $T_1(1324)$ as independent scalars.

Generalising this approach to two-index tensors, it turns out that it is sufficient to start with
\bea
T_{11}(1234) &=& (u_1 \gamma^m u_2)(u_3 \gamma^n  u_4) \,, \nn\\
T_{31}(1234) &=& (u_1 \gamma^a\gamma^m\gamma^n u_2)(u_3 \gamma_a  u_4) \,, \nn\\
T_{33}(1234) &=& (u_1 \gamma^{ab} \gamma^m u_2)(u_3 \gamma_{ab} \gamma^n  u_4) \,, \nn
\eea
and permutations of the spinor labels.  It would be very tiresome to systematically apply a variety of Fierz transformations by hand and to find an independent set.  Fortunately, by choosing a gamma matrix representation (such as the one listed in appendix \ref{s:gammamat}) and reducing all expressions to polynomials in the independent spinor components $u^1_i, \dots,  u^{16}_i$, this problem can be solved with computer help.  As expected, one finds that the $T_{ij}(abcd)$ span a nine-dimensional space, and a basis can be chosen as
\begin{multline}
\label{eq:ninetensors}
T_{11}(1234), T_{11}(1324), T_{11}(1423), T_{11}(3412), T_{11}(2413), T_{11}(2314), \\
 T_{31}(1234), T_{31}(1324), T_{31}(2314) \,.
\end{multline}
A typical relation reducing the other $T_{ij}(abcd)$ to this basis is
\be
\label{eq:4frelation1}
T_{31}(3412) = 2 T_{11}(1234) - 2 T_{11}(3412) + T_{31}(1324) + T_{31}(2314) + 2 \eta_{mn} T_1(1234) \,.
\ee

Having solved the first step, it is now easy to include the two or four momenta, taking the Dirac equation into account.  Consider first the case of two momenta.  Starting from the two-tensors in \eqref{eq:ninetensors}, one gets the three independent scalars
\be
(u_1\slashed{k}_3 u_2)(u_3\slashed{k}_1 u_4) \,, \qquad(u_1\slashed{k}_2 u_3)(u_2\slashed{k}_1 u_4) \,, \qquad
(u_1\slashed{k}_2 u_4)(u_2\slashed{k}_1 u_3) \,.
\nn
\ee
In addition, there are four products of the two independent scalars $T_1(1234)$ and $T_1(1324)$ with the two independent momentum invariants $\kk{1}{2}$ and $\kk{1}{3}$.  By contracting \eqref{eq:4frelation1} with momenta, one can show that
\begin{multline}
\label{eq:4frelation2}
\kk{1}{2} T_1(1324) -\kk{1}{3} T_1(1234) \\
= - (u_1\slashed{k}_3 u_2)(u_3\slashed{k}_1 u_4) + (u_1\slashed{k}_2 u_3)(u_2\slashed{k}_1 u_4) - (u_1\slashed{k}_2 u_4)(u_2\slashed{k}_1 u_3) \,,
\end{multline}
and this relation can be used to eliminate $\kk{1}{2} T_1(1324)$.  (It will become clear later that there are no independent other relations like this one.)  There are thus six independent $k^2 u_1\cdots u_4$ scalars:
\begin{align}
&(u_1\slashed{k}_3 u_2)(u_3\slashed{k}_1 u_4) && \kk{1}{2} \, T_1(1234) \nn\\
&(u_1\slashed{k}_2 u_3)(u_2\slashed{k}_1 u_4) && \kk{1}{3} \, T_1(1234) \label{eq:4fbasis6} \\
&(u_1\slashed{k}_2 u_4)(u_2\slashed{k}_1 u_3) && \kk{1}{3} \, T_1(1324) \nn
\end{align}
Note that there is only one completely antisymmetric combination of those, given by the right hand side of \eqref{eq:4frelation2}.  Similarly, in the case of four momenta, one finds ten independent $k^4 u_1\cdots u_4$ scalars:
\begin{align}
\basisf{1} &= \kk{1}{2} \, (u_1\slashed{k}_3 u_2)(u_3\slashed{k}_1 u_4) & &\basisf{2} = \kk{1}{3} \, (u_1\slashed{k}_3 u_2)(u_3\slashed{k}_1 u_4) \nn\\
\basisf{3} &= \kk{1}{2} \, (u_1\slashed{k}_2 u_3)(u_2\slashed{k}_1 u_4) & &\basisf{4} = \kk{1}{3} \, (u_1\slashed{k}_2 u_3)(u_2\slashed{k}_1 u_4) \nn \\
\basisf{5} &= \kk{1}{2} \, (u_1\slashed{k}_2 u_4)(u_2\slashed{k}_1 u_3) & &\basisf{6} = \kk{1}{3} \, (u_1\slashed{k}_2 u_4)(u_2\slashed{k}_1 u_3) \label{eq:4fbasis10} \\
\basisf{7} &= \kksq{1}{2} \, T_1(1234) & &\basisf{8} = \kk{1}{2}\kk{1}{3} \, T_1(1234) \nn\\
\basisf{9} &= \kksq{1}{3} \, T_1(1234) & &\basisf{10} = \kksq{1}{3} \, T_1(1324) \nn
\end{align}

Working in a gamma matrix representation, it is again simple to construct a computer algorithm which reduces any given $k^2 u_1\cdots u_4$ or $k^4 u_1\cdots u_4$ scalar into polynomials of the spinor and momentum components.  The Dirac equation can then be solved by breaking up the sixteen-component spinors $u_i$ into eight-dimensional chiral spinors $u_i^{\text{s}}$ and $u_i^{\text{c}}$, as in eq.~\eqref{eq:diracsoln}.  One obtains polynomials in the momentum components $k_i^a$ and the independent spinor components $(u_i^{\text{c}})^{1\dots 8}$.  However, a great disadvantage of this procedure is that it breaks manifest Lorentz invariance.  For example, one encounters expressions which contain subsets of terms proportional to the square of a single momentum and are therefore equal to zero, but it is difficult to recognise this with a simple algorithm.  The easiest solution is to choose several sets of particular vectors $k_i$ satisfying $k_i^2=0$ and $\sum_i k_i=0$ and to evaluate all expressions on these vectors.  (By choosing integer arithmetic, one easily avoids issues of numerical accuracy.)  Substituting these sets of momentum vectors in the bases \eqref{eq:4fbasis6} and \eqref{eq:4fbasis10} gives full rank six and ten respectively, showing they are indeed linearly independent.

Equipped with a computer algorithm for these basis decompositions, one finds, for example, that the symmetriser ${\cal S}_{\text{4F}}$ of section \ref{s:twoloop4f},
\bea
{\cal S}_{\text{4F}} = 4(1 + \pi_{12})(1-\pi_{23})(1+\pi_{24}) \,,
\nn
\eea
acts on the $\basisf{1}\dots \basisf{10}$ basis as
\bea
{\cal S}_{\text{4F}} \basisf{1} &=& -12 \basisf{4} + 12 \basisf{5} + 12 \basisf{6} \,, \nn \\
&\dots& \nn \\
{\cal S}_{\text{4F}} \basisf{10} &=& 8 \basisf{1} + 16 \basisf{2} - 8 \basisf{3} - 16 \basisf{4} + 8 \basisf{5} + 16 \basisf{6} -24 \basisf{7} - 24 \basisf{8} - 24 \basisf{9}
\nn
\eea
and has rank three.

\subsection{Two bosons, two fermions}
\label{s:basis2b2f}

The combined methods of the last two sections can easily be extended to the mixed case of two bosons and two fermions.  In the one-loop calculation of section \ref{s:oneloop2b2f}, one encounters scalars of the form $ku_1u_2F_3F_4$.  A basis of such objects is given by
\bea
\basisbf{1} &=& (u_1\gamma^a u_2) k^3_a F^3_{bc} F^4_{bc} \nn\\
\basisbf{2} &=& (u_1\gamma^a u_2) F^3_{ab} F^4_{bc} k^1_c \nn\\
\basisbf{3} &=& (u_1\gamma^a u_2) F^4_{ab} F^3_{bc} k^1_c \label{eq:2b2fbasis5}\\
\basisbf{4} &=& (u_1\gamma^{abc} u_2) F^3_{ab} F^4_{cd} k^3_d \nn\\
\basisbf{5} &=& (u_1\gamma^{abc} u_2) F^3_{ab} F^4_{cd} k^1_d \nn
\eea
There are two combinations antisymmetric in $[12]$ and symmetric in $(34)$:
\be
- \basisbf{1} + 4 \basisbf{2} + \basisbf{4} \qquad\text{and}\qquad \basisbf{2} + \basisbf{3} \,. \nn
\ee
Finally, there are ten independent scalars of the form $k^3u_1u_2F_3F_4$ (relevant to the two-loop calculation of section \ref{s:twoloop2b2f}), and they can all be obtained by multiplication of $\basisbf{1}\dots \basisbf{5}$ with the two momentum invariants $\kk{1}{2}$ and $\kk{1}{3}$.

\section{A gamma matrix representation}
\label{s:gammamat}

A convenient representation of the SO(1,9) gamma matrices is given by the $32 \times 32$ matrices
\be
\Gamma^a = \left(\begin{matrix} 0 & (\gamma^a)^{\alpha\beta} \\ (\gamma^a)_{\alpha\beta} & 0 \end{matrix}\right) \,,
\nn
\ee
where
\bea
(\gamma^0)^{\alpha\beta} &=& \mathbf{1}_{16} = (\gamma^0)_{\alpha\beta} \,, \nn\\
(\gamma^9)^{\alpha\beta} &=& \left(\begin{matrix} -\mathbf{1}_8 & 0 \\ 0 & \mathbf{1}_8 \end{matrix}\right) = - (\gamma^9)_{\alpha\beta} \,,
\nn
\eea
and $(\gamma^a)^{\alpha\beta} = - (\gamma^a)_{\alpha\beta}$, $a=1\dots 8$, is a real, symmetric $16 \times 16$ representation for the SO(8) Clifford algebra,
\be
(\gamma^a)^{\alpha\beta} = \left(\begin{matrix} 0 & \sigma^a \\ ({\sigma}^a)^T & 0 \end{matrix}\right)^{\alpha\beta} \,, \qquad a=1\dots 8 \,,
\nn
\ee
as given in appendix 5.B of \cite{gsw}.  The matrices $\Gamma^a$ satisfy the SO(1,9) Clifford algebra relations,
\be
{}\{\Gamma^a,\Gamma^b \} = 2 \eta^{ab} \mathbf{1}_{32} \,,\qquad \eta^{ab} = (+--\dots -) \,,
\nn
\ee
and bilinears of chiral spinors (with, say, positive chirality) are constructed as
\be
(u \Gamma^{[a_1\dots a_k]} v) = (u \gamma^{[a_1\dots a_k]} v) = u^\alpha (\gamma^{[a_1})_{\alpha\beta} (\gamma^{a_2})^{\beta\gamma} \dots (\gamma^{a_k]})_{\gamma\delta} v^\delta \,.
\nn
\ee
This representation is particularly suitable for the calculations outlined in appendix \ref{s:spinors} because it allows a simple decomposition of SO(1,9) spinors into SO(8) spinors due to its block structure:
\be
\Gamma^0 \cdots \Gamma^9 = \left(\begin{matrix} \mathbf{1}_{16} & 0 \\ 0 & -\mathbf{1}_{16} \end{matrix}\right) \,, \qquad \Gamma^1 \cdots \Gamma^8 = \left(\begin{matrix} \mathbf{1}_{8} & 0 & 0 & 0 \\ 0 & -\mathbf{1}_{8} & 0 & 0 \\ 0 & 0 & \mathbf{1}_{8} & 0 \\ 0 & 0 & 0 & -\mathbf{1}_{8} \end{matrix}\right)
\nn
\ee
Therefore, the Dirac equation for a chiral 16-component spinor $u$,
\be
(\gamma^a)_{\alpha\beta} \partial_a u^\alpha = 0 \,,
\nn
\ee
can be solved by splitting $u$ into two chiral eight-component spinors of SO(8),
\be
u = \left(\begin{matrix} u^{\text{s}} \\ u^{\text{c}} \end{matrix}\right) \qquad\text{with}\qquad \gamma^{1\dots 8} \left(\begin{matrix} u^{\text{s}} \\ u^{\text{c}} \end{matrix}\right)= \left(\begin{matrix} + u^{\text{s}} \\ - u^{\text{c}} \end{matrix}\right)\,.
\nn
\ee
One obtains the coupled equations
\bea
(\partial_0 + \partial_9) u^{\text{s}} - (\sigma\cdot\partial) u^{\text{c}} &=& 0 \nn\\
(\partial_0 - \partial_9) u^{\text{c}} - (\sigma^T\cdot\partial) u^{\text{s}} &=& 0 \nn
\eea
(with eight-dimensional dot products).  These can be solved for $u^{\text{s}}$ in terms of $u^{\text{c}}$:
\be
\label{eq:diracsoln}
u^{\text{s}} = \frac{-i}{\sqrt{2}k_+} (\sigma\cdot\partial) u^{\text{c}} = \frac{1}{\sqrt{2}k_+} (\sigma\cdot k) u^{\text{c}} \,,
\ee
where $k_+ = -i \partial_+ = \frac{- i}{\sqrt{2}} (\partial_0 + \partial_9)$.

\end{document}